\documentclass[1p]{elsarticle}
\usepackage{lineno}
\modulolinenumbers[5]
\usepackage[colorlinks=true]{hyperref}
\journal{---}

\usepackage{color}
\usepackage{amsmath}
\usepackage{amssymb}

\begin{document}

\begin{frontmatter}

\title{Explicit formulation of second and third order optical nonlinearity in the FDTD framework}

\author[ottawa,cegep]{Charles Varin\corref{correspondingauthor}}
\author[ottawa]{Rhys Emms}
\author[ottawa]{Graeme Bart}
\author[rostock]{Thomas Fennel}
\author[ottawa]{Thomas Brabec}

\cortext[correspondingauthor]{Corresponding author}

\address[ottawa]{Departement of physics, University of Ottawa, Ottawa (ON) K1N 6N5, Canada}
\address[cegep]{C\'egep de l'Outaouais, Gatineau (QC) J8Y 6M4, Canada}
\address[rostock]{Institut f\"{u}r Physik, Universit\"{a}t Rostock, 18051 Rostock, Germany}

\begin{abstract}
The finite-difference time-domain (FDTD) method is a flexible and powerful technique for rigorously solving Maxwell's equations. However, three-dimensional optical nonlinearity in current commercial and research FDTD softwares requires solving iteratively an implicit form of Maxwell's equations over the entire numerical space and at each time step. Reaching numerical convergence demands significant computational resources and practical implementation often requires major modifications to the core FDTD engine. In this paper, we present an explicit method to include second and third order optical nonlinearity in the FDTD framework based on a nonlinear generalization of the Lorentz dispersion model. A formal derivation of the nonlinear Lorentz dispersion equation is equally provided, starting from the quantum mechanical equations describing nonlinear optics in the two-level approximation. With the proposed approach, numerical integration of optical nonlinearity and dispersion in FDTD is intuitive, transparent, and fully explicit. A strong-field formulation is also proposed, which opens an interesting avenue for FDTD-based modelling of the extreme nonlinear optics phenomena involved in laser filamentation and femtosecond micromachining of dielectrics. 
\end{abstract}



\end{frontmatter}

\section{Introduction}
The finite-difference time-domain (FDTD) method is a powerful and flexible technique for the numerical integration of Maxwell's equations. It enables the rigorous study of optical, nanophotonic, and nanoplasmonic phenomena~\cite{taflove2005computational,taflove2013advances}. It uses centered finite differences applied to all electric and magnetic components of the electromagnetic field to achieve second-order integration accuracy. The algorithm originally proposed in 1966 by Yee~\cite{yee1966} solves the resulting set of coupled equations on three-dimensional numerical meshes, staggered in both space and time. Yee's approach remains today the most prominent numerical method for time-domain solutions of Maxwell's equations in complex settings. It is effectively used in commercial and research FDTD softwares~\cite{taflove2005computational,taflove2013advances} and is a key component of the particle-in-cell~\cite{Verboncoeur2005} and microscopic particle-in-cell~\cite{Varin2012,Peltz2012,Varin2014,Peltz2014} plasma simulation techniques.

As surprising as it seems, the inclusion of optical nonlinearity in the Yee-FDTD framework is not straightforward. It is easily demonstrated with the Amp\`ere's circuital law
\begin{align}\label{eq:ampere}
\nabla\times\mathbf{H} - \epsilon_0\frac{\partial\mathbf{E}}{\partial t} = \frac{\partial\mathbf{P}}{\partial t},
\end{align}
where the source term depends on a nonlinear susceptibility of the form
\begin{align} 
\mathbf{P} = \epsilon_0(\chi^{(1)}\mathbf{E} + \chi^{(2)}\mathbf{E}^2 + \chi^{(3)}\mathbf{E}^3+\ldots). 
\end{align}
This leads in Eq.~\eqref{eq:ampere} to a set of coupled nonlinear equations that are implicit in the electric field vector and whose solution is nontrivial~\cite{taflove2005computational,Goorjian1992}.
A formal solving approach, proposed by Greene and Taflove~\cite{taflove2013advances,Greene2006}, uses a recursive Newton method to obtain an approximate solution for $\mathbf{E}$. To ensure numerical convergence, it has to be performed over the entire numerical space at least a few times per time step. The Greene-Taflove iterative method is a rigorous way to include optical nonlinearity into the FDTD framework, but efficient implementation for solving three-dimensional problems is inherently complex. Typically, FDTD developers rely instead on explicit tricks whose implementation is simpler and computationally more efficient (see, e.g,~\cite{Maksymov2011}). But beyond the respective merits of the different implicit and explicit schemes, there is actually a need for transparent, physics-centric, nonlinear FDTD models that can provide a direct link between continuum optics and the quantum theory of optical polarizability. In this paper, we develop such a model based on the quantum two-level description of nonlinear optics.

As an alternative to the Greene-Taflove iterative technique, Gordon \emph{et al.} proposed the use of nonlinear oscillators whose integration in the Yee-FDTD framework is fully explicit and intuitive~\cite{Gordon2013}. Its connection to the Lorentz dispersion model---widely used to fit spectroscopic data---makes it a very appealing approach. However, time-domain integration of Gordon \emph{et al.}'s equations is unstable and should be used with care in time domain simulations like FDTD~\cite{VarinOE2015}. Nevertheless, we demonstrated previously that the inherent numerical instability of Gordon \emph{et al.}'s method is circumvented by introducing saturation to the nonlinearity term to mimic the dynamics of an atomic transition in the under-resonant limit~\cite{VarinOE2015}. However, the nonlinear oscillator equation presented in \cite{VarinOE2015} was introduced \emph{ad hoc}, without a formal mathematical proof. Moreover, in both \cite{Gordon2013} and \cite{VarinOE2015} the authors considered instantaneous nonlinearities only, while a rigorous treatment of nonlinear optics in dielectrics must include delayed contributions, in particular that associated with stimulated Raman scattering.

In this paper, we develop an approach for including optical nonlinearity in FDTD based on a nonlinear generalization of the Lorentz oscillator equation, commonly used to model linear optical dispersion (see, e.g.,~\cite{jackson1998electromagnetic}). It improves upon \cite{Gordon2013,VarinOE2015} by providing a formal derivation of the underlying oscillator equation, starting from the quantum mechanical equations for nonlinear optics in the two-level approximation, and by including a thorough discussion on how to deal with both the instantaneous (Kerr) and delayed (Raman) third-order nonlinearities. In particular, the proposed methodology is applied to the examples of second harmonic generation in periodic and plasmonic structures and to the propagation of intense femtosecond optical pulses in dielectrics. 

This paper is divided as follows. First in Sec.~\ref{sec:nonlinear}, we present the nonlinear Lorentz model and its explicit integration in the Yee-FDTD framework. A formal derivation from the quantum mechanical two-level model equations is given in~\ref{appendix:A}. Next, in Sec.~\ref{sec:example1} we apply the nonlinear Lorentz model to the modelling of second harmonic generation (SHG) in periodically poled Lithium Niobate (PPLN). In Sec.~\ref{sec:3D}, we look at the SHG enhancement in a dielectric by a split-ring resonator nano-antenna. In Sec.~\ref{sec:example2}, we present the nonlinear Lorentz model methodology to deal with short laser pulse propagation in  centrosymmetric dielectrics and apply it to the particular case of optical solitons. In Sec.~\ref{sec:strong}, we discuss a particular form of the nonlinear Lorentz model for strong field applications and use it to model self-focusing in a Kerr medium. We finally conclude in Sec.~\ref{sec:conclusions}.


\section{The nonlinear Lorentz model}
\label{sec:nonlinear}
Optical dispersion is typically modeled with the following ordinary differential equation~\cite{jackson1998electromagnetic,fowles1975introduction,boyd2008nonlinear}
\begin{align}\label{eq:lorentz_drude}
\frac{d^2 \mathbf{P}}{d \tau^2} + \frac{\gamma}{\omega_0} \frac{d \mathbf{P}}{d \tau} +  \mathbf{P} = \epsilon_0\bar{\chi}^{(1)} \mathbf{E},
\end{align}
also known as the Lorentz oscillator equation,
where $\gamma$ and $\omega_0$ are respectively the collision and resonance frequencies, $\tau = \omega_0 t$ is the oscillator proper time, $\epsilon_0$ is the vacuum permittivity, and $\bar{\chi}^{(1)}$ is the \emph{static} linear electric susceptibility. Eq.~\eqref{eq:lorentz_drude} is known to have the following solution in the Fourier domain
\begin{align}\label{eq:lorentz_drude_ft}
\tilde{\mathbf{P}}(\omega) = \epsilon_0\left(\frac{\omega_0^2}{\omega_0^2 - \omega^2 - i\gamma\omega}\right)\bar{\chi}^{(1)} \tilde{\mathbf{E}}(\omega).
\end{align}
Going from Eq.~\eqref{eq:lorentz_drude} to Eq.~\eqref{eq:lorentz_drude_ft}, harmonic oscillation of the electric field $\mathbf{E}$ and polarization density $\mathbf{P}$ at the angular frequency $\omega$ was assumed to be as $e^{-i\omega t}$. Eq.~\eqref{eq:lorentz_drude_ft} is used extensively to fit spectroscopic data, where multiple equations are combined to obtain satisfying dispersion curves over a certain spectral range, e.g., 
\begin{align}\label{eq:lorentz_drude_sum}
n^2(\omega) = 1 + \sum_{k}\left(\frac{\omega_k^2}{\omega_k^2 - \omega^2 - i\gamma_k\omega}\right)\bar{\chi}^{(1)}_k,
\end{align}
where the subscript $k$ refers to the parameters of the $k\mathrm{th}$ oscillator. In the particular case of dielectrics where conductivity can be neglected (i.e., all $\gamma_k \simeq 0$), Eq.~\eqref{eq:lorentz_drude_sum} takes a particular form known as the Sellmeier formula~\cite{fowles1975introduction,Sellmeier}.

The use of the oscillator proper time $\tau$ in Eq.~\eqref{eq:lorentz_drude} emphasizes the direct relationship between the static polarization density (i.e., $\mathbf{P}$ in the limit where $d \mathbf{P}/d \tau \sim 0$) and the linear driving term on the right-hand side ($\epsilon_0\bar{\chi}^{(1)} \mathbf{E}$). Intuitively, nonlinear contributions to the polarization density are introduced by adding nonlinear source terms, e.g.,
\begin{align}\label{eq:nonlinear_lorentz_drude}
\frac{d^2 \mathbf{P}}{d \tau^2} + \frac{\gamma}{\omega_0} \frac{d \mathbf{P}}{d \tau} +  \mathbf{P} = \epsilon_0\left(\bar{\chi}^{(1)} \mathbf{E}+\bar{\chi}^{(2)}\mathbf{E}^2 + \bar{\chi}^{(3)} \mathbf{E}^3+\ldots\right).
\end{align}
A benefit of this formulation compared with, e.g., the anharmonic oscillator model~\cite{Gordon2013,Scalora2015,boyd2008nonlinear}, is a direct correspondence between the different numerical susceptibility parameters $\bar{\chi}^{(n)}$ and those commonly used in perturbative nonlinear optics, generally obtained by measurements or \emph{ab initio} calculations~\cite{boyd2008nonlinear}. The formal derivation of Eq.~\eqref{eq:nonlinear_lorentz_drude} from the quantum mechanical two-level atom model in the under-resonant, adiabatic-following, weak-field limit is given in~\ref{appendix:A}.

In Eq.~\eqref{eq:nonlinear_lorentz_drude}, the right-hand side is associated with linear and nonlinear scattering, as well as harmonic generation, self-modulation, and wave-mixing processes, whereas the left-hand side defines how strongly the medium gets polarized by the different driving frequencies (dispersion). We stress that $\mathbf{E}$ is the total electric field vector, i.e., $\mathbf{E} = \mathbf{E}_0 + \mathbf{E}_1e^{-i\omega t} + \mathbf{E}_2e^{-i2\omega t} + \ldots$ It thus represents not only the incident signal oscillating at the dominant angular frequency $\omega_L$ of the laser spectrum, but also contributions coming from linear scattering ($\sim e^{-i\omega_L t}$) as well as nonlinear scattering and wave mixing ($\sim e^{-i0\omega_L t},e^{-i\omega_L t} e^{-i2\omega_L t},e^{-i3\omega_L t},\ldots$).

For simplicity, in this paper the optical susceptibilities and oscillator parameters in Eq.~\eqref{eq:nonlinear_lorentz_drude} are represented by scalars. However, a general formulation of the nonlinear Lorentz model must use tensors to account for the polarization-dependent nature of light-matter interactions and allow for proper modelling of the anisotropic optical response of a particular medium (see, e.g., Sec.~\ref{sec:3D}). Similarly to the Sellmeier's approach [see, e.g., Eq.~\eqref{eq:lorentz_drude_sum}], multiple nonlinear Lorentz oscillators can be used to model optical materials, to fit both the linear and nonlinear dispersion (see~\ref{appendix:dispersion}). The use of the nonlinear Lorentz dispersion model Eq.~\eqref{eq:nonlinear_lorentz_drude} in FDTD simulations is demonstrated with examples in Secs.~\ref{sec:example1}-\ref{sec:strong}.

\subsection{Explicit numerical integration in the FDTD framework}\label{sec:integration}
We now demonstrate that integration of the nonlinear Lorentz model in the FDTD framework can be made fully explicit. Following the standard Yee procedure to express Maxwell's equations in terms of finite differences~\cite{taflove2005computational}, one gets the following discretized equations for the electric and magnetic field vectors:
\begin{subequations}
\begin{align}\label{eq:maxwell_fd_1}
\mathbf{H}^{n+1/2} &=  \mathbf{H}^{n-1/2} - \frac{\Delta t}{\mu_0}
\left(\nabla\times\mathbf{E}\right)^{n},\\
\mathbf{E}^{n+1} &=  \mathbf{E}^{n} + \frac{\Delta t}{\epsilon_0}\left[
\left(\nabla\times\mathbf{H}\right)^{n+1/2} - \mathbf{J}^{n+1/2}\right],\label{eq:maxwell_fd_2}
\end{align}
\end{subequations}
where $n$ is the time index of the discretized time $t = n\Delta t$, with $\Delta t$ being the numerical time step. We omitted the spatial indices voluntarily for simplicity (for details see, e.g.,~\cite{taflove2005computational}). In this ``free space fields'' formulation, all material properties are contained in the current density $\mathbf{J}^{n+1/2}$. 

Using $d \mathbf{P}/d t = \mathbf{J}$, Eq.~\eqref{eq:nonlinear_lorentz_drude} is written in terms of two first-order ordinary differential equations whose discretization is done to match Yee's staggering scheme. This results in the following leapfrog integration equations for the material polarization and current densities:
\begin{subequations}
\begin{eqnarray}\label{eq:J_fd}
\mathbf{J}^{n + 1/2} &=& \frac{\left(1-\Gamma\right)}{
\left(1 + \Gamma\right)}\,\mathbf{J}^{n - 1/2} + \frac{\omega_0^2\Delta t}{
\left(1 + \Gamma\right)}\left(\bar{\mathbf{P}}^n - \mathbf{P}^n\right)\\
\mathbf{P}^{n + 1} &=&  \mathbf{P}^{n} + \Delta t\,\mathbf{J}^{n + 1/2}
\label{eq:P_fd}
\end{eqnarray}
\end{subequations}
where $\Gamma = \gamma\Delta t/2$ and $\bar{\mathbf{P}}^n = \epsilon_0[\bar{\chi}^{(1)}\mathbf{E}^n + \bar{\chi}^{(2)}(\mathbf{E}^n)^2 + \bar{\chi}^{(3)}(\mathbf{E}^n)^3+\ldots]$. It is seen immediately that in this form solving for $\mathbf{H}^{n+1/2}$ and $\mathbf{E}^{n+1}$, i.e., finding the electromagnetic field in the future, depends only on values in the past. Numerical integration of~Eqs.~(\ref{eq:maxwell_fd_1})-(\ref{eq:P_fd}) is thus fully explicit and does not require the Greene-Taflove iterative method~\cite{Greene2006}. 

In the next sections~\ref{sec:example1},~\ref{sec:3D},~\ref{sec:example2}, and~\ref{sec:strong}, we elaborate more on coupling the nonlinear Lorentz model to the FDTD technique by solving specific problems associated with light propagation in second and third order nonlinear media.

\section{Example 1: second-harmonic generation in a periodically poled material}
\label{sec:example1}

As a first example, we consider quasi-phase-matched second harmonic generation (QPM SHG) in a periodic structure, more specifically in periodically poled lithium niobate (PPLN)~\cite{Fejer1992,Byer1997,boyd2008nonlinear}. Technically, PPLN QPM SHG crystals are engineered by depositing metal electrodes directly on the crystal surface to which a strong voltage is applied to align the molecules and break the natural crystal symmetry~\cite{Byer1997}. By switching periodically the polarity of the strong voltage, it is possible to create a periodic array of domains with alternating positive and negative second order susceptibilities ($\pm\chi^{(2)}$). The reversal period is chosen to optimize the conversion efficiency of an input laser beam into its second harmonic. The modulation of the linear index is usually assumed to have a negligible impact~\cite{Fejer1992}. Below, we show how FDTD modelling of QPM SHG can be performed with the help of the nonlinear Lorentz equation presented above. 

\begin{figure}[tbp]
\centering
\includegraphics[width=0.75\columnwidth]{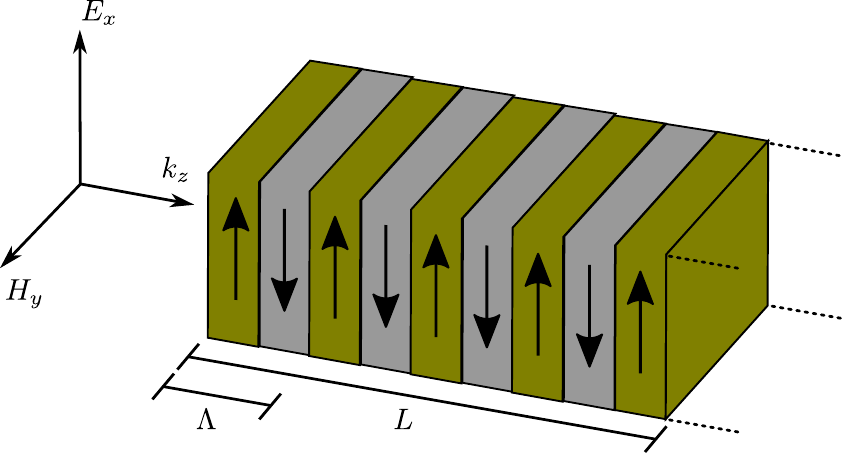}
\caption{Schematic representation of an electromagnetic plane wave impinging on an infinite PPLN crystal. Parameters $L$ and $\Lambda$ represent the distance away from the input face and the molecular orientation reversal period, respectively. The large-head black arrows indicate regions where the average permanent dipole moment is pointing up and down. With respect to the electric field $E_x$, this is associated with positive ($+\bar{\chi}^{(2)}_k$ for $\uparrow$) and negative ($-\bar{\chi}^{(2)}_k$ for $\downarrow$) second-order susceptibilities.}
\label{fig:SHG}
\end{figure}

For the demonstration, we considered a plane wave moving along the axis of an infinite PPLN crystal (see Fig.~\ref{fig:SHG}). For simplicity, we solved the corresponding 1D problem. To include both the linear and nonlinear responses of the periodically poled domains, we used polarization equations like the following:
\begin{align}\label{eq:2nd_order}
\frac{1}{\omega_k^2}\frac{d^2 P_k}{d t^2} +  P_k = \epsilon_0\left(\bar{\chi}^{(1)}_k E\pm \bar{\chi}^{(2)}_k E^2\right),
\end{align}
where the sign of $\bar{\chi}^{(2)}_k$ of the $k$th oscillator is chosen to account for the direction of the local average molecular alignment and associated average permanent dipole moment (see again Fig.~\ref{fig:SHG}). The linear refractive index is assumed to be constant throughout the medium. 

{\renewcommand{\arraystretch}{1.5}
\begin{table}[b]
\centering
\caption{Oscillator parameters used to model QPM SHG in periodically poled MgO:LN with a second-order nonlinear Lorentz equation [see Eq.~\eqref{eq:2nd_order}]. Typical effective second order nonlinearity $d_{\mathrm{eff}}$ of commercial QPM crystals is in the $14-17\,\mathrm{pm/V}$ range, corresponding to $\bar{\chi}^{(2)} \simeq 30\,\mathrm{pm/V}$.}\label{table:MGOLN}
\vspace{0.25cm}
\begin{tabular}{|c|c|c|c|}
\hline
$k$ & $\bar{\chi}^{(1)}_k$ & $\bar{\chi}^{(2)}_k$ (pm/V) & $\omega_k$ (rad/s)\\ \hline\hline
1 & 2.4272 & 30  & $1.5494\times 10^{16}$ \\ \hline
2 & 1.4617 & 0  & $7.9514\times 10^{15}$ \\ \hline
3 & 9.6536 & 0 & $9.7766\times 10^{13}$ \\
\hline
\end{tabular}
\end{table}

\begin{figure}[ht]
\centering
\includegraphics[width=0.6\columnwidth]{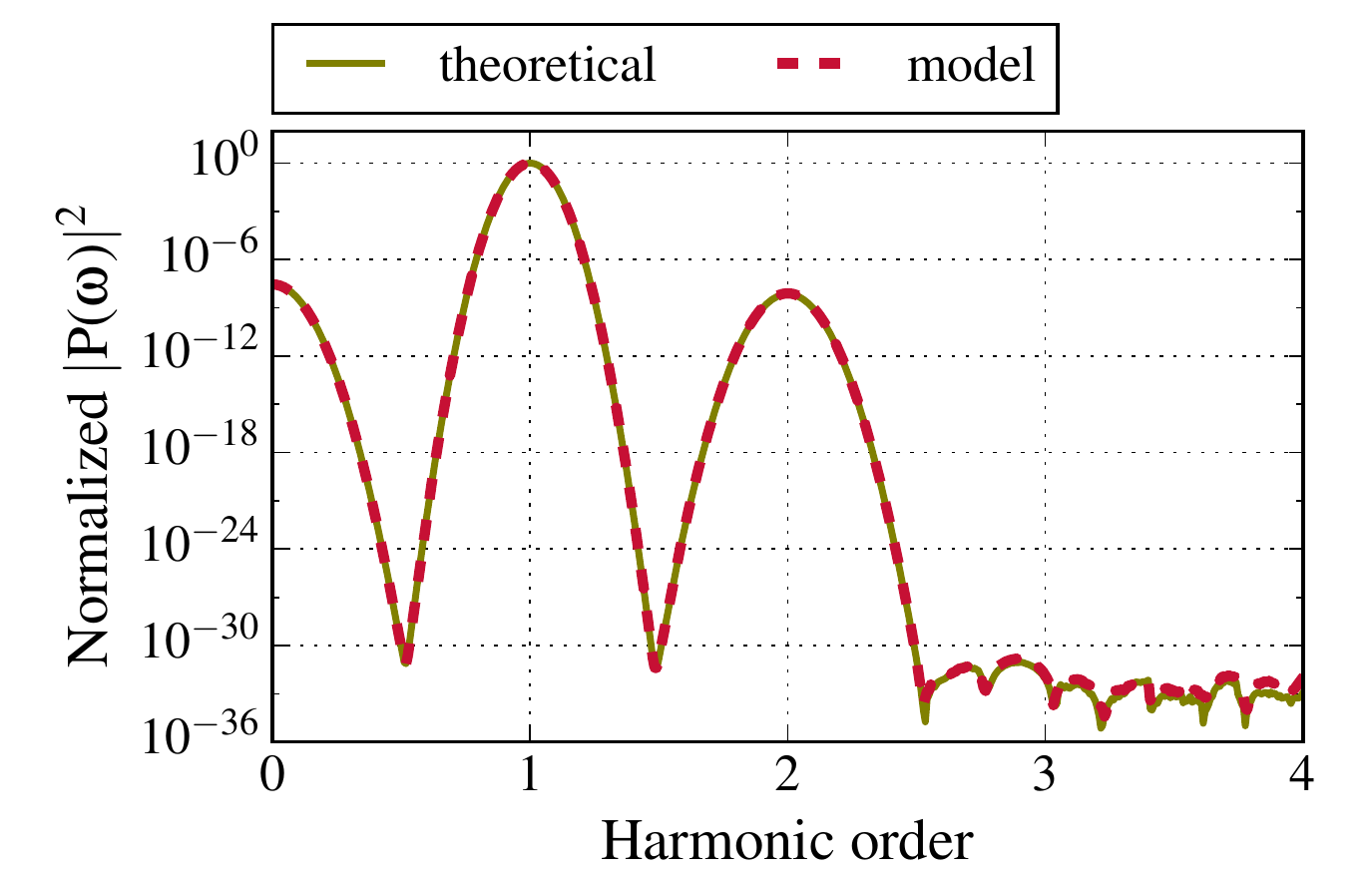}
\caption{Spectral response of the three-band model used for FDTD calculations of SHG in PPLN. Here we assumed a Gaussian pulse with central wavelength $\lambda_0=1.064\,\mu\mathrm{m}$, laser intensity $10^8\,\mathrm{W/cm}^2$, and pulse duration 20~fs. The dashed curve labeled ``model" refers to the explicit integration (Sec.~\ref{sec:integration}) of the three-band model described in the main text. The full ``theoretical" curve refers to a polarization given by $P(\omega) = \epsilon_0\mathrm{TF}\{\bar{\chi}^{(1)}E(t) + \bar{\chi}^{(2)}E^2(t)\}$, where $\mathrm{TF}\{\}$ is the Fourier transform and $\bar{\chi}^{(1)}\equiv n^2(2\pi c/\lambda_0) - 1$, as from Eq.~\eqref{eq:Sellmeier_LiNbO3}. The other parameters, including $\bar{\chi}^{(2)}$, are defined in Table~\ref{table:MGOLN}.}
\label{fig:LiNbO3}
\end{figure}

Effective modelling of optical dispersion in lithium niobate (LN) is achieved by using three oscillator equations like Eq.~\eqref{eq:2nd_order}. We chose the linear part to match the Sellmeier formula for MgO doped LN (MgO:LN)~\cite{ZelmonJOSAB97}:
\begin{align}\label{eq:Sellmeier_LiNbO3}
n^2(\omega) = 1 + \sum_{k=1}^3\left(\frac{\omega_k^2}{\omega_k^2 - \omega^2}\right)\bar{\chi}^{(1)}_k.
\end{align}
Values for the different parameters are given in Table \ref{table:MGOLN}. A typical value for the nonlinear susceptibility of commercial QPM crystals was added to one of the band equations according to Eq.~\eqref{eq:2nd_order}. The effective strength of the model second-order polarization depends on the relative position of the driving frequency with respect to the resonance of the band. The choice of a particular band was defined by numerical tests (see Fig.~\ref{fig:LiNbO3}). In principle, the nonlinearity can be added to any of the bands, and even split between them, as long as the static parameters are scaled accordingly to give a correct, total second-order polarization (see also~\ref{appendix:dispersion}).

In Fig.~\ref{fig:SHG_results}, we compared the nonlinear-Lorentz/FDTD analysis to a conventional theoretical model for QPM SHG (see \ref{appendix:shg}). For FDTD modelling, we performed simulations of a 40-$\mu$m-thick medium slab in a 
80-$\mu$m domain (ended by absorbing boundaries) discretized with a 4-nm resolution. A Gaussian laser pulse was initially outside the numerical domain and propagated through it for 0.8~ps with $\Delta t \simeq 13\,\mathrm{as}$. We sampled the electric field at 500 positions regularly spaced in the medium and took the absolute value of the Fourier transform of each of the 500 time traces [$E(t)$] to find the spectral amplitude of the second harmonic ($|E(2\omega_L)|$). These SH amplitudes were ultimately normalized by the amplitude of the incident electric field ($E_0 = \sqrt{2\eta_0 I}$, with $I$ being the laser intensity and $\eta_0$ the characteristic impedance of vacuum) and mapped to the corresponding position in the medium. Results are shown in Fig.~\ref{fig:SHG_results}. Using higher spatial and temporal resolution did not change the results significantly.

\begin{figure}[ht]
\centering
\includegraphics[width=0.6\columnwidth]{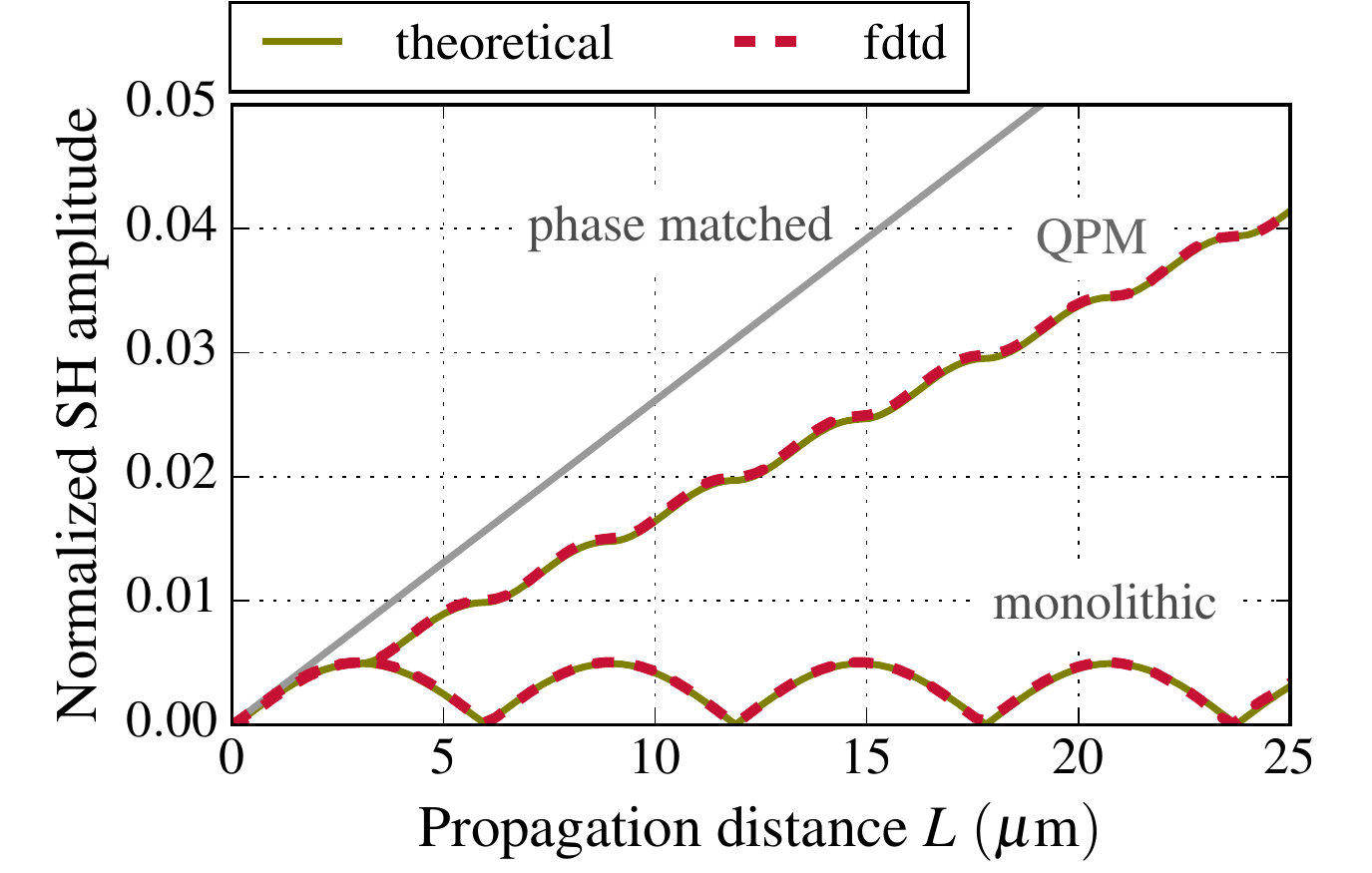}
\caption{Nonlinear-Lorentz/FDTD analysis of SHG in a QPM PPLN crystal. The field amplitude of the second harmonic signal as a function of the distance $L$ is compared with theory (see \ref{appendix:shg}). Laser pulse parameters were: intensity $5\times 10^{8}\,\mathrm{W/cm}^2$, duration 10 fs, and central wavelength $\lambda_0 = 1.064\,\mu\mathrm{m}$. To account for optical reflection at the input crystal face and different definitions of the field amplitude in both models, the intensity used in the theoretical model was rescaled to match the SH growth rate predicted by FDTD. The difference between the two models is barely noticeable.}
\label{fig:SHG_results}
\end{figure}

For a medium without a periodic structure (monolithic), the amplitude of the second-harmonic signal ($2\omega_L$) oscillates due to the wave vector mismatch with respect to the input signal ($\omega_L$). The oscillation period $\Lambda$ depends effectively on the mismatch parameter $\Delta k = (2\omega_L/c) [n(\omega_L) - n(2\omega_L)]$, while the amplitude of the oscillation is proportional to $\bar{\chi}^{(2)}I$~\cite{boyd2008nonlinear}. The excellent agreement between theory and FDTD in Fig.~\ref{fig:SHG_results} demonstrates that the nonlinear-Lorentz/FDTD analysis succeeds in reproducing quantitatively the dispersion, scattering, and wave mixing processes in SHG, as well as their interplay.

Quasi phase matching (QPM) consists in bypassing the wave vector mismatch by inverting the sign of $\chi^{(2)}$ periodically to make the SH signal grow continuously. With the parameters presented in Table~\ref{table:MGOLN}, the optimal sign reversal period is $\Lambda_{\mathrm{QPM,\, opt}} = 2\pi/\Delta k\simeq 5.9~\mu\mathrm{m}$. Periodic poling was implemented in both the theoretical model and FDTD by setting the sign of $\bar{\chi}^{(2)}$ with the spatial function $\mathrm{sign}[\sin(2\pi z/\Lambda_{\mathrm{QPM,\, opt}})]$, with $z=0$ corresponding to the input face of the crystal. The agreement between FDTD and theory for the QPM curves in Fig.~\ref{fig:SHG_results} demonstrates in addition the ability of the nonlinear-Lorentz/FDTD analysis to model linear and nonlinear optical processes correctly in the presence of a spatial modulation of the refractive index.

Summing up, in this section we have shown that quantitative insight into subwavelength nonlinear scattering and wave-mixing processes is possible with the nonlinear-Lorentz/FDTD approach. A three-dimensional generalization of the one-dimensional analysis we have provided is straightforward. Finite transverse beam profile and medium response anisotropy (birefringence) are easy to include as well. The flexibility and reliability of the nonlinear-Lorentz/FDTD method offer an exceptional potential for modelling SHG and other nonlinear phenomena in complex settings, e.g., for the analysis of SHG in metametarial devices~\cite{Lepetit2015a,Segal2015}. In the next section, we develop further on this topic by modelling in three dimensions the enhancement of SHG in the vicinity of a nanometric plasmonic structure.

\section{Example 2: SHG enhancement by a split-ring resonator}
\label{sec:3D}
In the previous section, we demonstrated the potential of the nonlinear-Lorentz/FDTD approach to model second-order optical processes in monolithic and periodic structures. To solve a one-dimensional formulation of the quasi-phase-matching problem, we used an effective, scalar nonlinear parameter $d_{\mathrm{eff}}$ to characterize the interaction between the incoming and scattered light with the nonlinear medium. However, for rigorous modelling of three-dimensional (3D) optics in solids, it is often necessary to consider the tensorial nature of both the linear and nonlinear susceptibilities. Below, we use the nonlinear-Lorentz dispersion model to perform 3D vectorial FDTD simulations of the SHG enhancement by a split-ring resonator (SRR) nano-antenna.

The nonlinear conversion of radiation into its second harmonic (SH) from a periodic array of SRRs~(see, e.g.,~\cite{Linden2012,Segal2015,OBrien2015}) is a representative application of plasmonic metamaterials. SRR arrays typically consist in gold nano-SRRs with thickness in the few-tens-of-nm range deposited on an optically transparent substrate (see, e.g.,~\cite{Niesler2009,Niesler2011,Linden2012}). The origin of the enhanced SHG emission has been much debated, and effectively depends on the specific alignment of the host (substrate) crystal planes with the gap of the SRR nano-antenna. Electromagnetic FDTD modelling is widely used for the design and optimization of SRR nano-antennas and to analyze experimental measurements.

FDTD modelling of SRR arrays is typically done by simulating a single SRR nano-antenna in a numerical domain surrounded by periodic boundaries in the directions parallel to the substrate surface to mimic an SRR array extending to infinity~(see, in particular,~\cite{Linden2012}). The 3D geometry of the computer model used here closely follows that found in~\cite{Niesler2009} and is shown in Fig.~\ref{fig:3D_SRR}. To define the linear dielectric constants of the nano-antenna (gold, Au) and the substrate (gallium arsenide, GaAs), we fit experimental data with oscillator equations in order to obtain reasonable refractive indexes and extinction coefficients at both the wavelength of the impinging laser pulse ($\lambda_L=1.5\,\mu\mathrm{m}$) and its second harmonic ($750\,\mathrm{nm}$). Details are given in Fig.~\ref{fig:GaAs_Gold} and below. 

\begin{figure}[t]
	\centering
	\includegraphics[width=0.8\columnwidth]{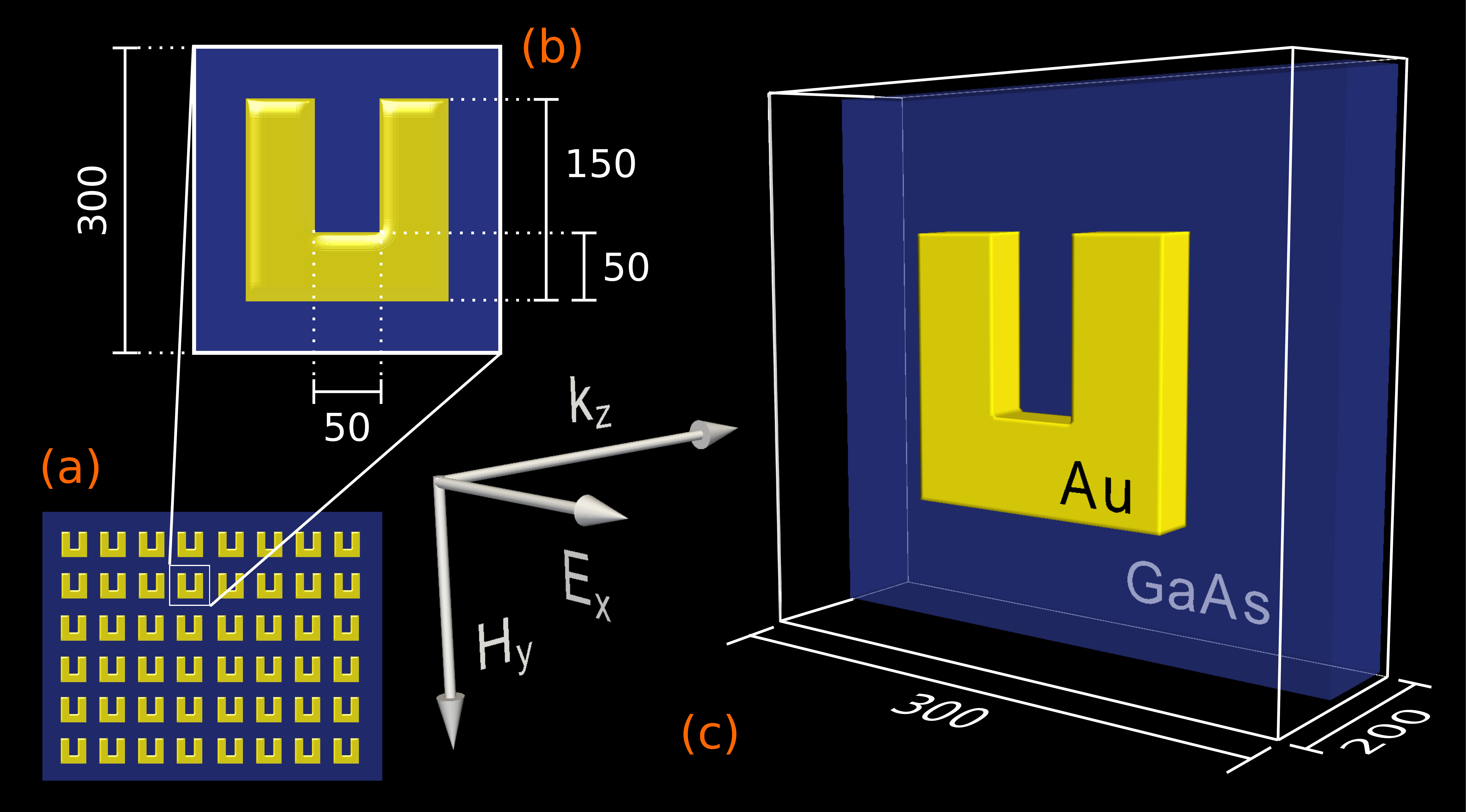}
	\caption{Schematic representation of the 3D FDTD SRR simulation. To model the 2D array of gold (Au) SRRs nano-antennas deposited on a gallium arsenide substrate (GaAs) shown in (a), a single SRR unit cell in (b) is modelled with periodic boundary conditions in the $x$ and $y$ directions, and absorbing boundaries in $z$ (the direction of light propagation). The 3D simulation layout is given in (c), with the axis system showing the direction of the impinging plane wave ($k_z$) whose electric field is linearly polarized along the $x$ direction. All measures are in nanometers (nm). For the simulations we performed, the thickness of the gold SRR and the GaAs substrate is 30~nm~\cite{Linden2012}. The particular SRR shown in (b) has a resonance around $1.5~\mu\mathrm{m}$~\cite{BuschPR2007}.} 
	\label{fig:3D_SRR}.
\end{figure}

For simplicity, we model here the gold optical response with a linear Drude polarization model, expressed below as a second-order, damped-oscillator equation: 
\begin{equation}\label{eq:Drude}
\frac{d^2\mathbf{P}_\mathrm{Au}}{dt^2} +\gamma_p\frac{d\mathbf{P}_\mathrm{Au}}{dt} = \epsilon_0\omega_p^2\mathbf{E},
\end{equation}
where $\omega_p$ and $\gamma_p$ are the plasma and collision frequencies, respectively. Fit values for gold are given in Fig.~\ref{fig:GaAs_Gold}(b) (see also~\cite{EtchgoinJCP2007erratum}). It is readily observed that Eq.~\eqref{eq:Drude} has the same structure as Eq.~\eqref{eq:nonlinear_lorentz_drude} and that it integrates also explicitly in FDTD using the method presented in Sec.~\ref{sec:integration}.

Gallium arsenide (GaAs) is an interesting medium for the current demonstration. Its zincblende crystal structure possesses a cubic lattice that does not display linear birefringence and has a fairly simple second-order nonlinear dielectric susceptibility tensor~(see, e.g., \cite{boyd2008nonlinear}). Assuming the same crystal-SRR orientation as in~\cite{Niesler2009}, the static second-order polarization components in the transverse $x-y$ plane read:
\begin{subequations}\label{eq:P2_GaAs_static}
\begin{align}
	P_x^{(2)} & = \epsilon_0\bar{\chi}^{(2)}\left(E_zE_z - E_yE_y\right),\\	
	P_y^{(2)} & = -2\epsilon_0\bar{\chi}^{(2)}E_xE_y.
\end{align}
\end{subequations}
Combining these two with the oscillator equation obtained from the fit of the experimental data for the linear dispersion [see Fig.~\ref{fig:GaAs_Gold}(a)], we can write a second-order nonlinear-Lorentz equation as follows:
\begin{align}\label{eq:GaAs}
\frac{1}{\omega_0^2}\frac{d^2\mathbf{P}_{\rm GaAs}}{dt^2} +\frac{\gamma_0}{\omega_0^2}\frac{d\mathbf{P}_{\rm GaAs}}{dt} + \mathbf{P}_{\rm GaAs} = &\epsilon_0\bar{\chi}^{(1)}\mathbf{E}\nonumber\\ &+ \epsilon_0\bar{\chi}^{(2)}\left[\left(E_zE_z - E_yE_y\right)\mathbf{i} - 2E_xE_y\,\mathbf{j}\right],
\end{align}
where $\mathbf{i}$ and $\mathbf{j}$ are unit vectors along the $x$ and $y$ coordinate axes, respectively. Values for the linear parameters are found in the caption of Fig.~\ref{fig:GaAs_Gold}. The second-order parameter was set to the accepted value of 
$\bar{\chi}^{(2)} = 7.4\times 10^{-10}\,\mathrm{m/V}$~\cite{boyd2008nonlinear}. 

For the sake of simplicity, we have modeled here linear and nonlinear dispersion in one equation. In a more realistic scenario, different oscillator equations should be used to model accurately the frequency dependence of both the linear and nonlinear responses. More information is provided in \ref{appendix:dispersion}. 


\begin{figure}[t]
	\centering
	\includegraphics[width=1.0\columnwidth]{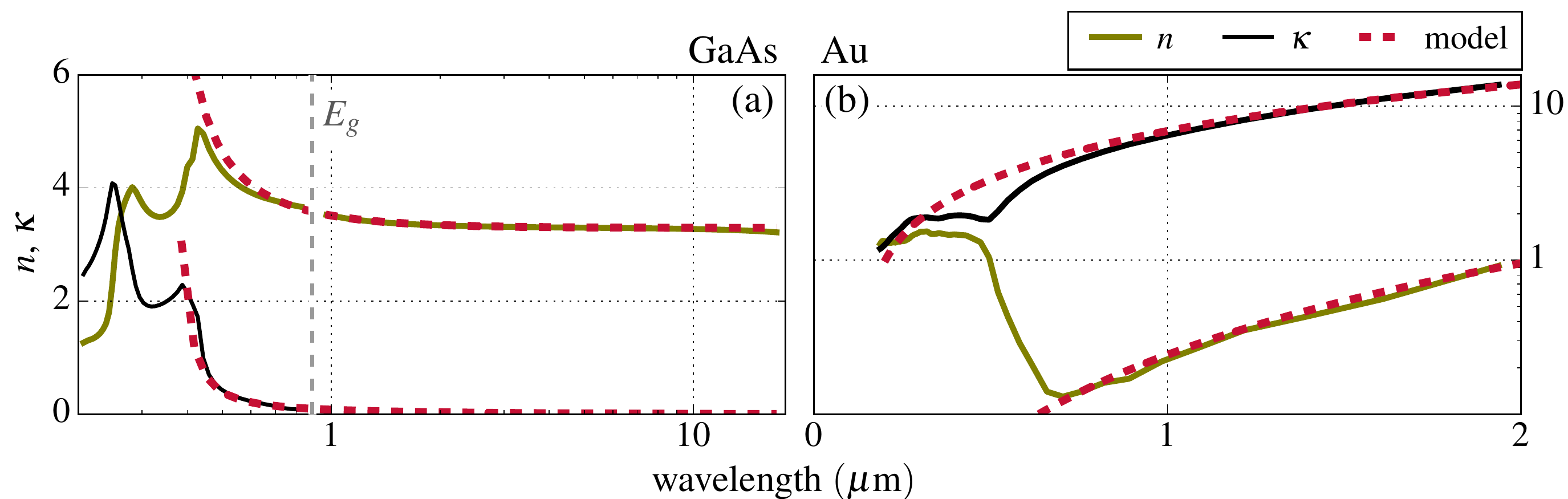}
	\caption{Linear optical dispersion in (a) gallium arsenide (GaAs) and (b) gold (Au). Both the refractive index ($n = \mathrm{Re}\{\sqrt{\epsilon}\}$) and extinction coefficient ($\kappa = \mathrm{Im}\{\sqrt{\epsilon}\}$) are shown as solid lines and compared with the simplified numerical model (red dashed lines). Experimental data in (a) come from~\cite{ApsnesJAP1986,JellisonOM1992} (for photon energy above the band gap energy $E_g\simeq 1.4\,\mathrm{eV}$) and \cite{SkauliJAP2003} (below $E_g$). In (b), experimental data come from~\cite{JohnsonPRB1972}. Numerical model fit in (a) corresponds to $\epsilon_{\rm GaAs} = 1 + \bar{\chi}^{(1)}\omega_0^2/(\omega_0^2 - \omega^2 - i\gamma_0\omega)$ with $\bar{\chi}^{(1)} = 9.85$, $\omega_0 = 5.18\times 10^{15}\,\mathrm{rad/s}$, and $\gamma_0 = 6\times 10^{14}\,\mathrm{rad/s}$. Model fit in (b) corresponds to the Drude contribution given in~\cite{EtchgoinJCP2007erratum}, more specifically to $\epsilon_{\rm Au} = 1 - \omega_p^2/(\omega^2 + i\gamma_p\omega)$, with $\omega_p = 1.317\times 10^{16}\,\mathrm{rad/s}$ and $\gamma_p = 1.3\times 10^{14}\,\mathrm{rad/s}$. For integration in FDTD, both $\epsilon_{\rm GaAs}$ and $\epsilon_{\rm Au}$ are converted into time-domain equations by taking the inverse Fourier transform of the corresponding spectral-domain polarization density $\mathbf{P}(\omega) = [\epsilon(\omega) - 1]\mathbf{E}(\omega)$ (for further detail, see main text). In both cases, discretization of the resulting second-order differential equation is done as in Sec.~\ref{sec:integration}.} 
	\label{fig:GaAs_Gold}.
\end{figure}

Inspection of the material models given above indicate that second harmonic should not be 
generated in either the gold SRR nano-antenna or the GaAs substrate alone by a plane wave with its electric field polarized along $x$. This is confirmed in FDTD simulations with each of the materials considered separately [see Fig.~\ref{fig:SHG_SRR}(a)]. However, when both media are present, SHG emerges [see, again, Fig.~\ref{fig:SHG_SRR}(a)]. This is in agreement with the results reported in~\cite{Niesler2009}. Inspection of Eq.~\eqref{eq:GaAs} suggests that the presence of the second harmonic in $x$ arises from the $y$ and $z$ components of the electric field around the nano-antenna.

The spectral analysis of the average polarization density of gold and GaAs shows that even and odd harmonics are present in both media [see, again, Fig.~\ref{fig:SHG_SRR}(b)], although the gold polarization model is linear (and cannot produce harmonics on its own). This reveals a wave mixing effect that we interpret as follows. 1) The laser field at $\omega$ induces a linear plasmonic field around the gold SRR that drives SHG in GaAs. 2) The SH radiation from GaAs goes into the gold SRR and creates a second-order nanoplasmonic field. 3)~That second-order nanoplasmonic field acts back onto the GaAs substrate, where it can induce a linear polarization at $2\omega$ and mix with other frequencies through $\chi^{(2)}$. 4)~The radiation field from that first interaction loop can propagate into the SRR to induce various orders of nanoplasmonic fields, and the loop continues. Such complex coupling effects are hard to account for in a purely analytical treatment, but is fully embedded in FDTD simulations. 

With this second example, we have shown that the nonlinear Lorentz model allows intuitive and flexible 3D FDTD modelling of nanophotonics and nanoplasmonics devices composed of several materials, including dielectrics with an anisotropic susceptibility tensor. We emphasize that current approaches to model the nonlinearity from metal nano-antennas are based on the classical Maxwell-Vlasov theory~\cite{ZengPRB2009,LiuJCP2010}, that gives a more detailed account of the metal plasma response. 


\begin{figure}[t]
	\centering
	\includegraphics[width=1.0\columnwidth]{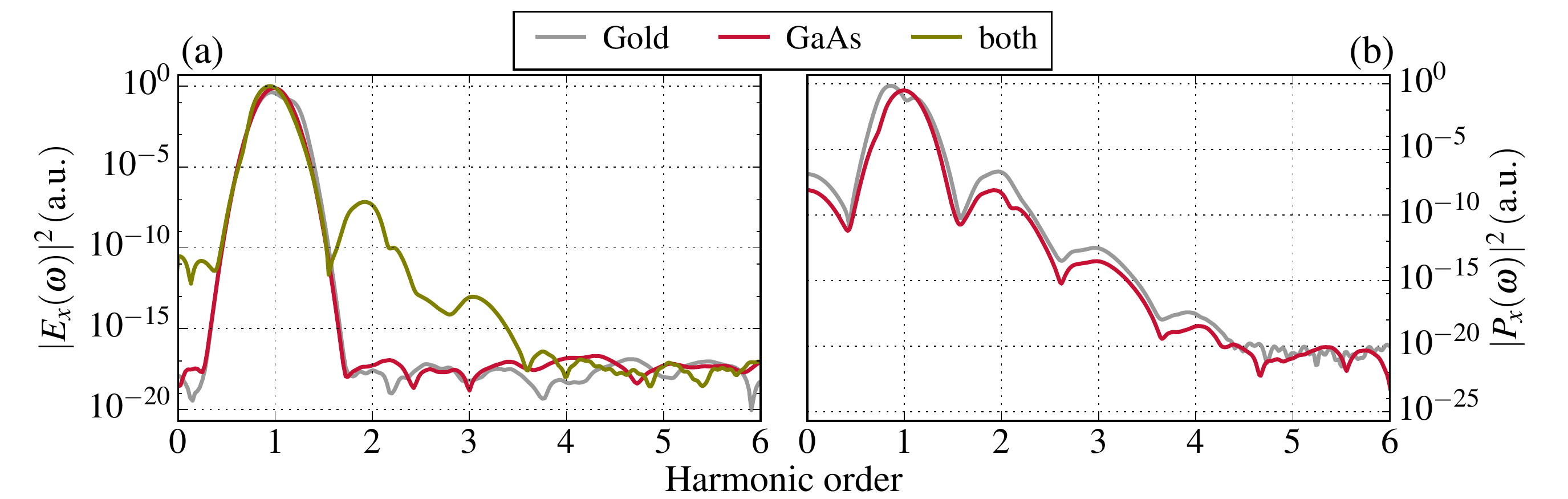}
	\caption{Spectral analysis of the 3D FDTD model described in Fig.~\ref{fig:3D_SRR}. In (a), the spectrum of the $x$ component of the radiated electric field recorded after the device indicates that SHG arises only when both media are present. With the current model, SHG from the GaAs substrate is triggered by the inhomogeneous field around the nano-antenna. The graph in (b) reveals the interplay between the two media that results in a complex 
    wave-mixing process that is naturally accounted for by the nonlinear-Lorentz/FDTD model.} 
	\label{fig:SHG_SRR}.
\end{figure}

\section{Example 3: short pulse propagation in centrosymmetric dielectrics}
\label{sec:example2}

In centrosymmetric dielectrics, the third order nonlinearity is responsible for third harmonic generation and the intensity-dependent modification of the refractive index (the optical Kerr effect). For laser pulses with a finite duration, an accurate description of these processes must include both the nearly-instantaneous electronic response (instantaneous Kerr) and a delayed component associated with stimulated molecular Raman scattering (Raman)~\cite{boyd2008nonlinear}. This is suitably represented by the following equation for the third-order polarization density~\cite{Agrawal}:
\begin{align}\label{eq:3rd_order_response}
\mathbf{P}^{(3)}(t) = \epsilon_0\chi^{(3)}\mathbf{E}\int_{-\infty}^t g(t - t')\left|\mathbf{E}(t') \right|^2dt',
\end{align}
where $g(t) = \alpha\delta(t) + (1-\alpha) h_{R}(t)$. Good approximate forms for the instantaneous Kerr and Raman response functions are, respectively, a Dirac delta function $\delta(t)$ and
\begin{align}\label{eq:Raman_function}
h_R(t) = \left( \frac{\tau_1^2 + \tau_2^2}{\tau_1\tau_2^2} \right) e^{-t/\tau_2}\sin(t/\tau_1),
\end{align}
with parameters $\tau_1$ and $\tau_2$ chosen to fit the Raman-gain spectrum~\cite{Agrawal}. 
The balance between the instantaneous and delayed contributions is parametrized by $\alpha$. Eq.~\eqref{eq:3rd_order_response} is often written in terms of a time-dependent rotation/vibration coordinate $Q$ as
\begin{align}\label{eq:3rd_order_current_Greene}
\mathbf{P}^{(3)}(t) = \underbrace{\epsilon_0\chi^{(3)}\alpha\mathbf{E}^3}_{\text{instantaneous Kerr}} + \underbrace{\epsilon_0\chi^{(3)}(1 - \alpha) Q\mathbf{E}}_{\text{Raman}},
\end{align}
where
\begin{align}\label{eq:Q_convolution}
Q(t) = \int_{-\infty}^t h_R(t - t')\left|\mathbf{E}(t') \right|^2dt'.
\end{align}

Effectively, Eq.~\eqref{eq:Raman_function} is the retarded Green's function of a damped harmonic oscillator with resonance angular frequency $\omega_R = \sqrt{(\tau_1^2 + \tau_2^2)/\tau_1^2\tau_2^2}$ and damping constant  $\gamma_R = 1/\tau_2$. Then, an alternative equation to Eq.~\eqref{eq:Q_convolution} for the generalized rotation/vibration coordinate $Q$ is~(see also~\cite{Greene2006,Couairon2011,Kolesik2014})
\begin{align}\label{eq:Q}
\frac{d^2 Q}{dt^2} + 2\gamma_R\frac{d Q}{dt} + \omega_R^2 Q = \omega_R^2 \left|\mathbf{E}\right|^2.
\end{align}
Below, we use this differential form to avoid to compute the convolution.

For the integration of Eqs.~(\ref{eq:3rd_order_current_Greene}) and~(\ref{eq:Q}) in FDTD, Greene and Taflove~\cite{Greene2006,taflove2013advances} proposed to define a current density from two consecutive polarization densities, e.g., 
\begin{align}\label{eq:3rd_order_response_no_conv}
\mathbf{J}^{n+1/2}_{Raman} \propto \frac{Q^{n+1}\mathbf{E}^{n+1} - Q^{n}\mathbf{E}^{n}}{\Delta t},
\end{align}
which leads to implicit FDTD equations that need to be solved iteratively to obtain an approximate solution for $\mathbf{E}^{n+1}$~\cite{taflove2013advances,Greene2006}. Instead, explicit FDTD integration with delayed third-order nonlinearity can be done by replacing $\epsilon_0\bar{\chi}^{(3)}\mathbf{E}^3$ in Eq.~\eqref{eq:nonlinear_lorentz_drude} by the right-hand side of equation Eq.~\eqref{eq:3rd_order_current_Greene} to give the following nonlinear Lorentz equation:
\begin{align}\label{eq:nonlinear_3rd_order}
\frac{d^2 \mathbf{P}}{d \tau^2} + \frac{\gamma}{\omega_0} \frac{d \mathbf{P}}{d \tau} +  \mathbf{P} = \epsilon_0\bar{\chi}^{(1)} \mathbf{E} + \epsilon_0\bar{\chi}^{(3)}\left[\alpha\mathbf{E}^3 + (1 - \alpha) Q\mathbf{E}\right],
\end{align}
where the evolution of $Q$ is given by Eq.~\eqref{eq:Q}. 

With $G = dQ/dt$, expressing Eqs.~(\ref{eq:Q}) and~(\ref{eq:nonlinear_3rd_order}) in terms of finite differences leads to the fully-explicit 4-step FDTD update sequence that follows:
\begin{enumerate}
\item Evaluate $\bar{\mathbf{P}}^n = \epsilon_0\bar{\chi}^{(1)} \mathbf{E}^n + \epsilon_0\bar{\chi}^{(3)}\left[\alpha (\mathbf{E}^n)^3 + (1 - \alpha) Q^n\mathbf{E}^n\right]$.
\item Update the current and polarization densities with Eqs.~(\ref{eq:J_fd})~and~(\ref{eq:P_fd}).
\item Update $Q$ with
\begin{subequations}
\begin{eqnarray}\label{eq:G_fd}
G^{n + 1/2} &=& \frac{\left(1-\Gamma_R\right)}{
\left(1 + \Gamma_R\right)}\,G^{n - 1/2} + \frac{\omega_R^2\Delta t}{
\left(1 + \Gamma_R\right)}\left[(\mathbf{E}^n)^2 - Q^n\right]\\
Q^{n + 1} &=&  Q^{n} + \Delta t\,G^{n + 1/2},
\label{eq:Q_fd}
\end{eqnarray}
\end{subequations}
where $\Gamma_R = \Delta t \gamma_R$.
\item Update the electromagnetic field with Eqs.~(\ref{eq:maxwell_fd_1}) and~(\ref{eq:maxwell_fd_2}).
\end{enumerate}

As a practical example, we considered the 1D propagation of a short optical pulse in fused silica and used a moving numerical window to track the pulse (the window moves with the pulse at the group velocity)~\cite{Arber2015}. Medium parameters are summarized in Table~\ref{table:Si02} (see table caption for other definitions used below). 

\begin{table*}[htbp]
\centering
\caption{Parameters used to model optical pulse propagation in fused silica~\cite{Agrawal,MALITSON1965}. The third order susceptibility parameter $\bar{\chi}^{(3)}_k$ was obtained from the third order nonlinear index $n_2 = 2.59\times10^{-20}~\mathrm{m^2/W}$ measured at a reference wavelength of 1.5~$\mu\mathrm{m}$~\cite{Agrawal}. The corresponding nonlinear length is $L_\mathrm{NL} = (n_2I\omega_0/c)^{-1}\simeq 9.22\times 10^{12}~\mathrm{m}/I$, where $I$ is the laser intensity expressed in $\mathrm{W/m^2}$. The group velocity dispersion (GVD) parameter $\beta_2 \simeq -22.2~\mathrm{fs}^2/\mathrm{mm}$ is readily obtained from the Sellmeier equation, which allows to define the dispersion length $L_D = T^2/|\beta_2|\simeq 45~\mu\mathrm{m}\times (T[\mathrm{fs}])^2$, where $T[\mathrm{fs}]$ is the pulse duration in femtosecond.
}\label{table:Si02}
\vspace{0.25cm}
\begin{tabular}{|c|c|c|c|c|c|c|c|}
\hline
$k$ & $\bar{\chi}^{(1)}_k$ & $\omega_k$ (rad/s) & $\bar{\chi}^{(3)}_k$ (m$^2$/V$^2$) & $\alpha$ & $\omega_R$ (rad/s) &  $\gamma_R$ (1/s)\\ \hline\hline
1 & 0.69617 & $2.7537\times 10^{16}$ & $1.94\times 10^{-22}$ & 0.7 &  $8.7722\times 10^{13}$  & $3.1250\times 10^{13}$\\ \hline
2 & 0.40794 & $1.6205\times 10^{16}$ & 0  & 0   & 0  & 0\\ \hline
3 & 0.89748 & $1.9034\times 10^{14}$ & 0  & 0   & 0  & 0\\
\hline
\end{tabular}
\end{table*}

We recall that the propagation of intense and short light pulses in dielectrics is directly influenced by both the linear and nonlinear responses of the medium. We have shown in the previous examples that these effects are correctly accounted for by the nonlinear Lorentz model. In 1D and in the presence of anomalous dispersion ($\beta_2 < 0$), the interplay between group velocity dispersion (GVD) and self-phase modulation (SPM) can manifest itself in a spectacular way by the formation of temporal optical solitons~\cite{Agrawal}. In the case of a fundamental soliton ($N^2 = L_D/L_\mathrm{NL} \sim 1$), GVD and SPM are perfectly balanced and the pulse shape and spectrum remain constant during propagation. This particular optical effect was used by several authors to test nonlinear models in FDTD~\cite{Greene2006,Fujii2004,Goorjian1992b}. For a 10-fs pulse at a 1.5-$\mu\mathrm{m}$ wavelength in fused silica, the fundamental soliton condition is fulfilled when the laser intensity is on the order of $10^{11}~\mathrm{W/cm}^2$. Results shown in Fig.~\ref{fig:propagation} are in good agreement with this prediction and similar tests found in~\cite{Greene2006,Fujii2004,Goorjian1992b}. This demonstrates, again, the capability of the nonlinear-Lorentz/FDTD approach to model quantitatively linear and nonlinear optical processes and their interplay.

\begin{figure*}[ht]
\centering
\includegraphics[width=\columnwidth]{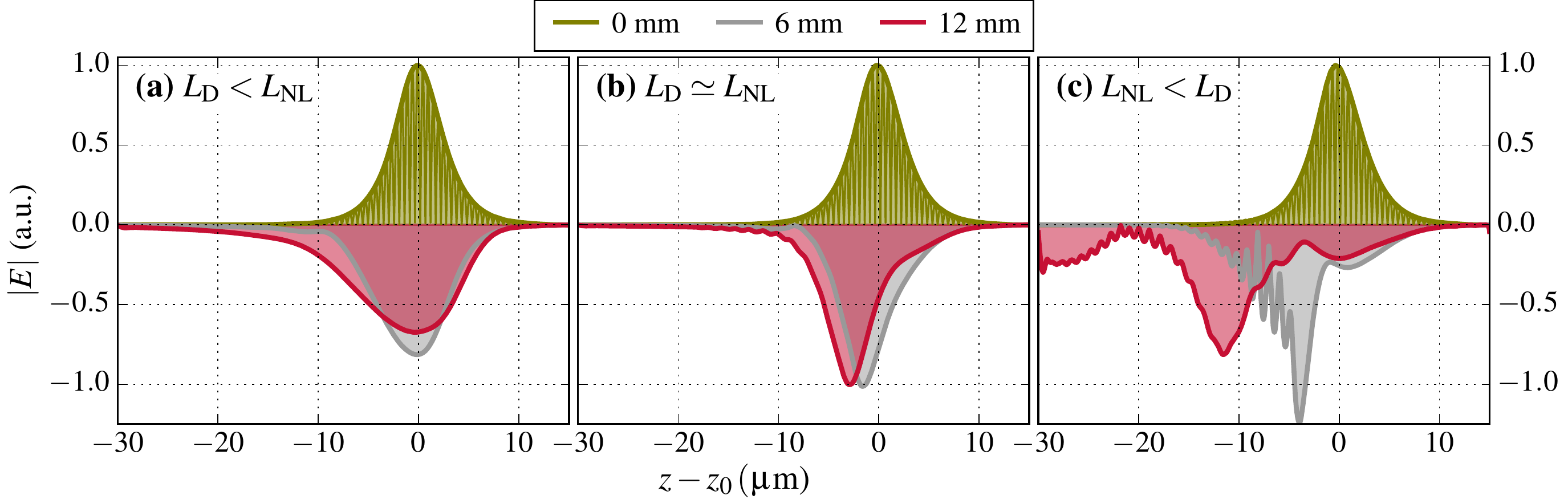}
\caption{Nonlinear-Lorentz/FDTD analysis of soliton propagation in fused silica. Color-coded are the pulse envelopes obtained by taking the absolute value of the Hilbert transform of the pulse snapshots at different propagation distances $z_0$ (see legend). In (a) where $I = 10^{8}\,\mathrm{W/cm}^2$, group velocity dispersion dominates and the initial pulse [$\propto\mathrm{sech}(t/t_0)$, with $t_0 = 10\,\mathrm{fs}$] spreads as it propagates. In (b) where $I = 3.25\times 10^{11}\,\mathrm{W/cm}^2$, the pulse propagates with nearly-constant width and amplitude over a distance greater than the soliton period $\pi L_D/2\simeq 7\,\mathrm{mm}$, suggesting that it is effectively a fundamental soliton~\cite{Agrawal}. The shift of the peak toward the left [$(z-z_0) < 0$] indicates that this soliton propagates at a slower group velocity than the pulse in (a). In (c) where $I = 10^{12}\,\mathrm{W/cm}^2$, nonlinearity dominates and the pulse self-phase modulates temporally before breaking up due to modulation instability. We stress that the origin of the $z$ axis, i.e., $z = z_0$, corresponds to a reference point moving at the linear group velocity.
}
\label{fig:propagation}
\end{figure*}

We emphasize that in previous works~\cite{Greene2006,Fujii2004,Goorjian1992b}, simulations were done with scaled parameters to ensure that optical solitons were observable within short propagation distances ($< 200~\mu\mathrm{m}$) with reasonable computational resources. In particular in~\cite{Fujii2004}, authors used $\chi^{(3)} = 7\times 10^{-2}\,\mathrm{m^2}/\mathrm{V^2}$ with a 1 V/m pulse peak electric field amplitude. With physically relevant material parameters like in the current work, this is equivalent to a laser intensity of $\sim 4.8\times 10^{13}\,\mathrm{W/cm}^2$, which is typically the light intensity inside a laser filament~\cite{Berge2007,couaironPR2007}. In this extreme regime the nonlinear material response is non-perturbative and accurate modelling of light-matter interaction must take material ionization and breakdown into account. Proper coverage of this complex topic is far beyond the scope of this paper. Nevertheless, in the next section we propose a particular form of the nonlinear Lorentz model suitable for strong-field applications, and test it in the context of self-focusing of a femtosecond laser pulse in a gas.

\section{Example 4: self-focusing in a dielectric}
\label{sec:strong}

For this last example, we consider self-focusing of an intense optical pulse in a dielectric due to the intensity-dependent modification of the refractive index associated with the third order nonlinearity. This phenomenon, where an optical pulse collapses onto itself, is typically associated with strong electric fields, with peak amplitudes beyond material breakdown thresholds. For both solids and gases, it is accepted that Kerr self-focusing is usually counter-acted by a rapid plasma build-up associated with a self-defocusing effect. In turn, laser filaments emerge when Kerr self-focusing and plasma defocusing balance each other~\cite{Berge2007,couaironPR2007}.

A rigorous study of laser filamentation is beyond the scope of this paper. However, there are still open questions related to laser filament dynamics and how to efficiently control the self-organization process~\cite{Berge2007,couaironPR2007}. Some of these questions are directly related to general optical phenomena like dispersion, third and fifth harmonic generation, the vectorial nature of light, and light propagation beyond the paraxial and slowly-varying-envelope approximations. All of which can be only fully accounted for by solving self-consistently Maxwell's equations. With the example provided below, we introduce a strong field version of the nonlinear-Lorentz/FDTD approach that can be used to explore this extreme phenomenon, provided that it would be complemented with proper models for field ionization and free-carrier dynamics.

In Table~\ref{table:strongfield}, we compiled different oscillator equations to integrate explicitly the optical Kerr nonlinearity in FDTD via Eqs.~(\ref{eq:J_fd}) and~(\ref{eq:P_fd}). In the under-resonant ($\omega_L \ll \omega_0$) weak-field ($\bar{\chi}^{(3)}|\mathbf{E}|^2\ll \bar{\chi}^{(1)}$) limit, they all trivially converge to the Lorentz dispersion model of Eq.~\eqref{eq:lorentz_drude_ft}. But we will see that their respective strong-field behaviour is significantly different. Their behaviour with respect to the driving frequency is addressed in~\ref{appendix:validity}.

{\renewcommand{\arraystretch}{1.5}
	\begin{table}[h]
		\centering
		\caption{Comparison between different nonlinear harmonic oscillator models that follow an equation like $\frac{d^2 \mathbf{P}}{d \tau^2} + \frac{\gamma}{\omega_0} \frac{d \mathbf{P}}{d \tau} +  \mathbf{P} = \bar{\mathbf{P}},$ where $\bar{\mathbf{P}}$ is a static polarization density that contains the nonlinearity and $\tau = \omega_0t$ is the oscillator proper time. The parameter $b$ below is the anharmonic constant, proportional to an effective $\chi^{(3)}$ (see \cite{Scalora2015,boyd2008nonlinear,Gordon2013} and \ref{appendix:A}). All these oscillator models can be integrated explicitly in FDTD via Eqs.~(\ref{eq:J_fd}) and~(\ref{eq:P_fd}), with the appropriate form for $\bar{\mathbf{P}}$. A numerical comparison is given in Figs.~\ref{fig:strong_field} and~\ref{fig:strong_field_I_scan} (see also~\ref{appendix:validity} for further analysis).}\label{table:strongfield}
		\vspace{0.25cm}
		\begin{tabular}{|p{2.0cm}|c|c|c|}
			\hline
			Name & $\bar{\mathbf{P}}$ & Refs.\\ \hline\hline
			Lorentz & $\epsilon_0\bar{\chi}^{(1)}\mathbf{E}$ &  \cite{fowles1975introduction}  \\ \hline
			Nonlinear Lorentz & $\epsilon_0\left(\bar{\chi}^{(1)} + \bar{\chi}^{(3)}|\mathbf{E}|^2\right)\mathbf{E}$ & Eq.~\eqref{eq:nonlinear_lorentz_drude}    \\ \hline
			Anharmonic & $\epsilon_0\bar{\chi}^{(1)}\mathbf{E} + b\mathbf{P}^3$ & \cite{Gordon2013,boyd2008nonlinear,Scalora2015}, Eq.~\eqref{eq:anharmonic_oscillator}\\ \hline
			Strong-field & see Eqs.~(\ref{eq:nonlinear_lorentz_sat}) and~(\ref{eq:nonlinear_lorentz_sat2}) & see also~\cite{VarinOE2015} \\ 
			\hline
		\end{tabular}
	\end{table}

\begin{figure*}[ht]
	\centering
	\includegraphics[width=\columnwidth]{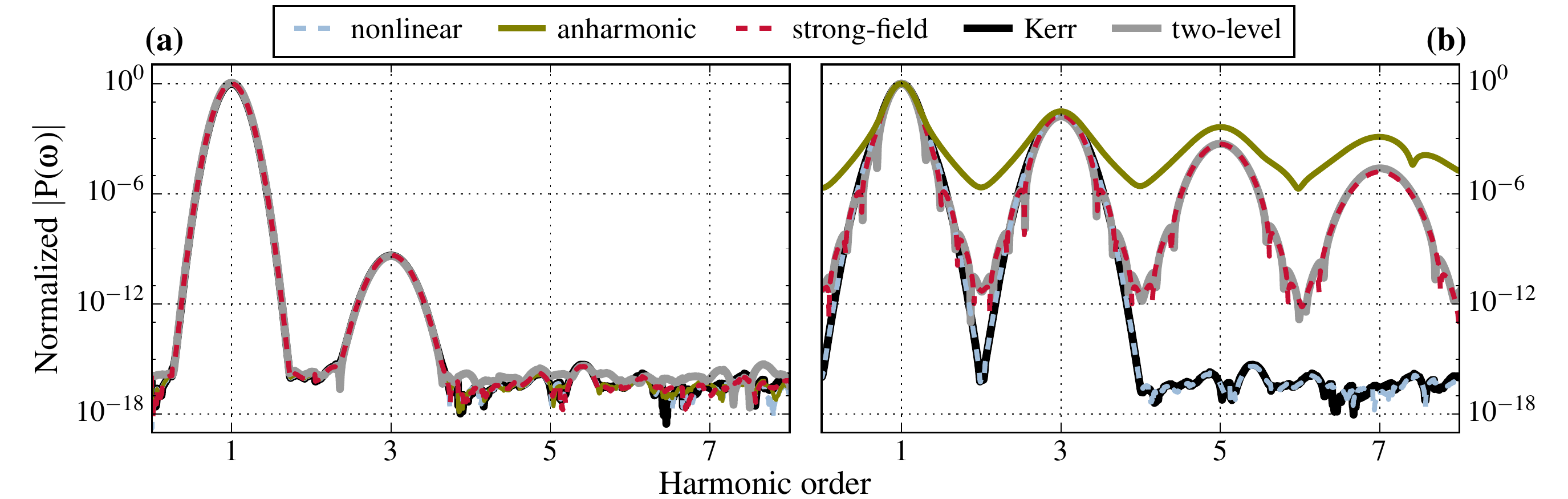}
	\caption{Comparison between the nonlinear oscillator models summarized in Table~\ref{table:strongfield}. While in (a) their weak-field ($I = 10^6\,\mathrm{W/cm}^2$) solutions match perfectly, in (b) their strong-field behaviors ($I = 5\times 10^{13}\,\mathrm{W/cm}^2$) differ significantly. In (b), the strong-field Lorentz model characterized by Eqs.~(\ref{eq:nonlinear_lorentz_sat}) and~(\ref{eq:nonlinear_lorentz_sat2}) can effectively reproduce the polarization of the quantum mechanical two-level model (see~\ref{appendix:schrodinger}), whereas the nonlinear Lorentz model [see Eq.~\eqref{eq:nonlinear_lorentz_drude}] follows the instantaneous Kerr polarization $P = \epsilon_0\left(\bar{\chi}^{(1)}E + \bar{\chi}^{(3)}E^3\right)$. For strong fields, the integration of the anharmonic oscillator is unstable and fails at reproducing any of the two-level or instantaneous Kerr polarizations. For these tests, we used fictitious values for the oscillator parameters ($\gamma = 0$ and $\omega_0 = 3\times 10^{16}\,\mathrm{rad/s}$) and susceptibilities ($\bar{\chi}^{(1)} = 5\times 10^{-4}$ and $\bar{\chi}^{(3)} = 1.6\times 10^{-25}\,\mathrm{m}^2/\mathrm{V}^2$) similar to those for air at ambient temperature and pressure. The two-level model and the anaharmonic $b$ parameters were set so that all results agree in the weak-field limit shown in (a). The driving electric field was that of a 10-fs Gaussian pulse with $\lambda_L = 800\,\mathrm{nm}$.}
	\label{fig:strong_field}
\end{figure*}

\begin{figure*}[ht]
	\centering
	\includegraphics[width=\columnwidth]{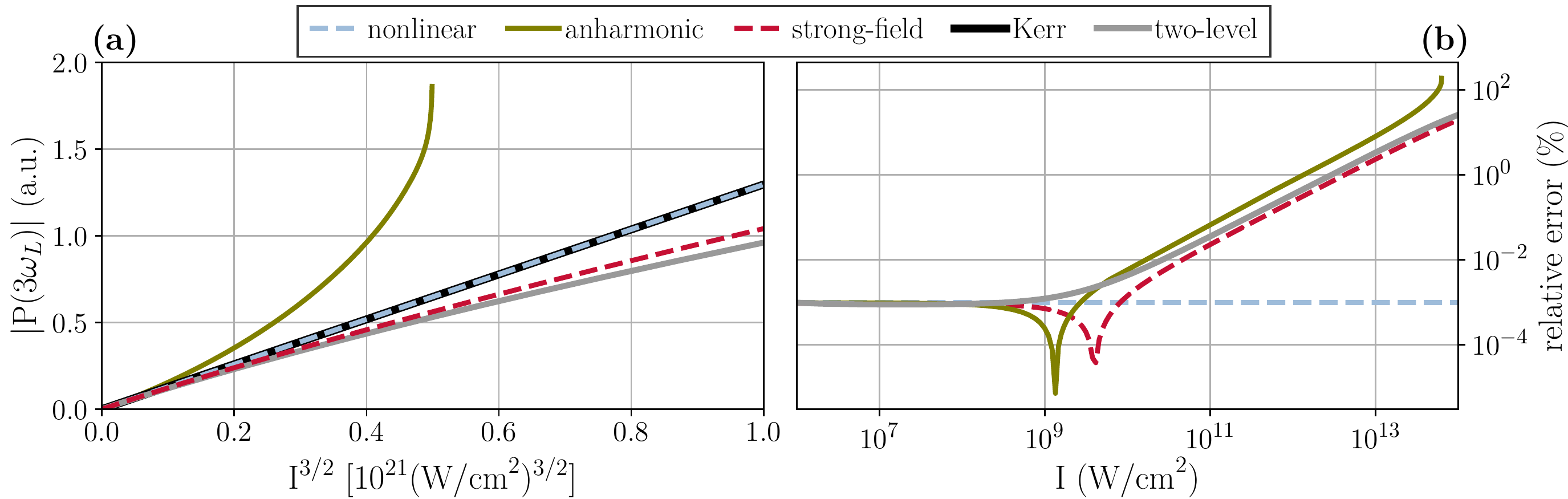}
	\caption{Intensity-scaling analysis of the third-harmonic polarization amplitude obtained with the nonlinear harmonic oscillator models given in Table~\ref{table:strongfield} (see also caption of Fig.~\ref{fig:strong_field} for details). In (a), plotting $|P(3\omega)|$ as a function of $I^{3/2}$ emphasizes the three following things : 1. the perfect agreement between the nonlinear Lorentz and the instantaneous Kerr models, 2. the appearance of optical saturation in both the strong-field and quantum two-level models, and 3. the numerical instability of the anharmonic model. In (b), we plotted the relative error to the Kerr model for each of the other models. The relative error is defined as $100\times|P_{3,model} - P_{3,Kerr}|/P_{3,Kerr}$, where $P_{3,Kerr} = \epsilon_0\bar{\chi}^{(3)}E_0^3 = \epsilon_0\bar{\chi}^{(3)} (2\eta_0 )^{3/2}I^{3/2}$, and $P_{3,model}$ is given by the amplitude of the third-harmonic polarization $P(3\omega_L)$ for each of the numerical models (nonlinear, anharmonic, strong-field, and quantum).
	}
	\label{fig:strong_field_I_scan}
\end{figure*}

A rigorous treatment of strong-field optical phenomena requires quantum mechanics~(see, e.g.,~\cite{KolesikOptica2014}). Nevertheless, in a previous work~\cite{VarinOE2015} we have shown that the driving force of an oscillator can be chosen in a way to reproduce the polarization of a quantum mechanical two-level atom in the under-resonant limit. In terms of the formalism developed in the current paper, the strong-field oscillator of~\cite{VarinOE2015} reads
\begin{align}\label{eq:nonlinear_lorentz_sat}
\frac{d^2 \mathbf{P}}{d \tau^2} + \frac{\gamma}{\omega_0} \frac{d \mathbf{P}}{d \tau} +  \mathbf{P} = \epsilon_0\bar{\chi}^{(1)}\left(\frac{1}{1+\bar{\chi}^{(3)}|\mathbf{E}|^2/\bar{\chi}^{(1)}}\right)\mathbf{E}.
\end{align}
In the weak-field limit, it expands to
\begin{align}\label{eq:nonlinear_lorentz_sat_taylor}
\frac{d^2 \mathbf{P}}{d \tau^2} + \frac{\gamma}{\omega_0} \frac{d \mathbf{P}}{d \tau} +  \mathbf{P} = \epsilon_0\left(\bar{\chi}^{(1)}-\bar{\chi}^{(3)}|\mathbf{E}|^2 + \ldots\right)\mathbf{E}.
\end{align}
Comparing Eq.~\eqref{eq:nonlinear_lorentz_sat_taylor} with Eq.~\eqref{eq:nonlinear_lorentz_drude}, it appears that Eq.~\eqref{eq:nonlinear_lorentz_sat} effectively models the polarizability of a centrosymmetric material ($\bar{\chi}^{(2)}=0$) with a \emph{negative} Kerr nonlinearity. It is straightforward to deduce from the expected weak-field expansion for a \emph{positive} Kerr effect, i.e.,
\begin{align}\label{eq:nonlinear_lorentz_sat2_taylor}
\frac{d^2 \mathbf{P}}{d \tau^2} + \frac{\gamma}{\omega_0} \frac{d \mathbf{P}}{d \tau} +  \mathbf{P} = \epsilon_0\left(\bar{\chi}^{(1)}+\bar{\chi}^{(3)}|\mathbf{E}|^2 - \ldots\right)\mathbf{E},
\end{align}
that the corresponding strong-field oscillator equation should be 
\begin{align}\label{eq:nonlinear_lorentz_sat2}
\frac{d^2 \mathbf{P}}{d \tau^2} + \frac{\gamma}{\omega_0} \frac{d \mathbf{P}}{d \tau} +  \mathbf{P} = \epsilon_0\bar{\chi}^{(1)}\left(2- \frac{1}{1+\bar{\chi}^{(3)}|\mathbf{E}|^2/\bar{\chi}^{(1)}}\right)\mathbf{E}.
\end{align}
The specific form of Eq.~\eqref{eq:nonlinear_lorentz_sat2} is here obtained \emph{ad hoc}, however it is in agreement with the formal solution presented in~\ref{appendix:A} where it is assumed that the population of the two-level atom is entirely in the ground state in the absence of a driving electric field [see, e.g., Eq.~\eqref{eq:nonlinear_lorentz}].

Although Eqs.~(\ref{eq:nonlinear_lorentz_sat}) and~(\ref{eq:nonlinear_lorentz_sat2}) are characterized by identical linear and nonlinear polarization spectral amplitudes $|\tilde{\mathbf{P}}(\omega)|$, they differ by the relative phase between the linear and the third harmonic signals. This difference is such that a laser beam propagating in a medium modelled with Eq.~(\ref{eq:nonlinear_lorentz_sat}) will experience self-defocusing ($-\bar{\chi}^{(3)}$), whereas in a medium modelled with Eq.~(\ref{eq:nonlinear_lorentz_sat2}) it will experience self-focusing ($\bar{\chi}^{(3)}$). 


We compared the numerical behaviour of the models given in Table~\ref{table:strongfield} against the quantum mechanical two-level model given in \ref{appendix:schrodinger}. In the strong-field limit, the results in Fig.~\ref{fig:strong_field} clearly show that only the strong-field Lorentz model characterized by Eqs.~(\ref{eq:nonlinear_lorentz_sat}) and~(\ref{eq:nonlinear_lorentz_sat2}) can effectively reproduce the response associated with the quantum mechanical two-level equation, whereas the nonlinear Lorentz model follows an instantaneous Kerr polarization of the form $P = \epsilon_0\left(\bar{\chi}^{(1)}E + \bar{\chi}^{(3)}E^3\right)$. 

Deeper insight was gained by plotting the amplitude of the third-harmonic polarization $|P(3\omega_L)|$ as a function of the laser intensity $I=E_0^2/2\eta_0$, where $\omega_L = 2\pi c/\lambda_L$ is the laser central laser frequency. In particular, we compared the numerically integrated models with the I-scaling behaviour expected from an instantaneous Kerr polarization, i.e.,  $P(3\omega_L) = \epsilon_0\bar{\chi}^{(3)}E_0^3 = \epsilon_0\bar{\chi}^{(3)} (2\eta_0 )^{3/2}I^{3/2}$, which is linear if plotted as a function of $I^{3/2}$. The results given in Fig.~\ref{fig:strong_field_I_scan} further emphasize the two observations of Fig.~\ref{fig:strong_field} as follows. 1. The nonlinear Lorentz model reproduces the instantaneous Kerr behaviour over the entire range of intensity considered with a relative error in the $10^{-5}$ range. 2. The strong-field model offers a reasonable approximation to the quantum two-level model. We recall that there is here an exceptional agreement between the dynamic nonlinear Lorentz and instantaneous Kerr models because simulations are performed in the under-resonant limit. The range of validity of the nonlinear Lorentz model with respect to the driving (laser) frequency is addressed in \ref{appendix:validity}.

\begin{figure*}[]
\centering
\includegraphics[width=0.6\columnwidth]{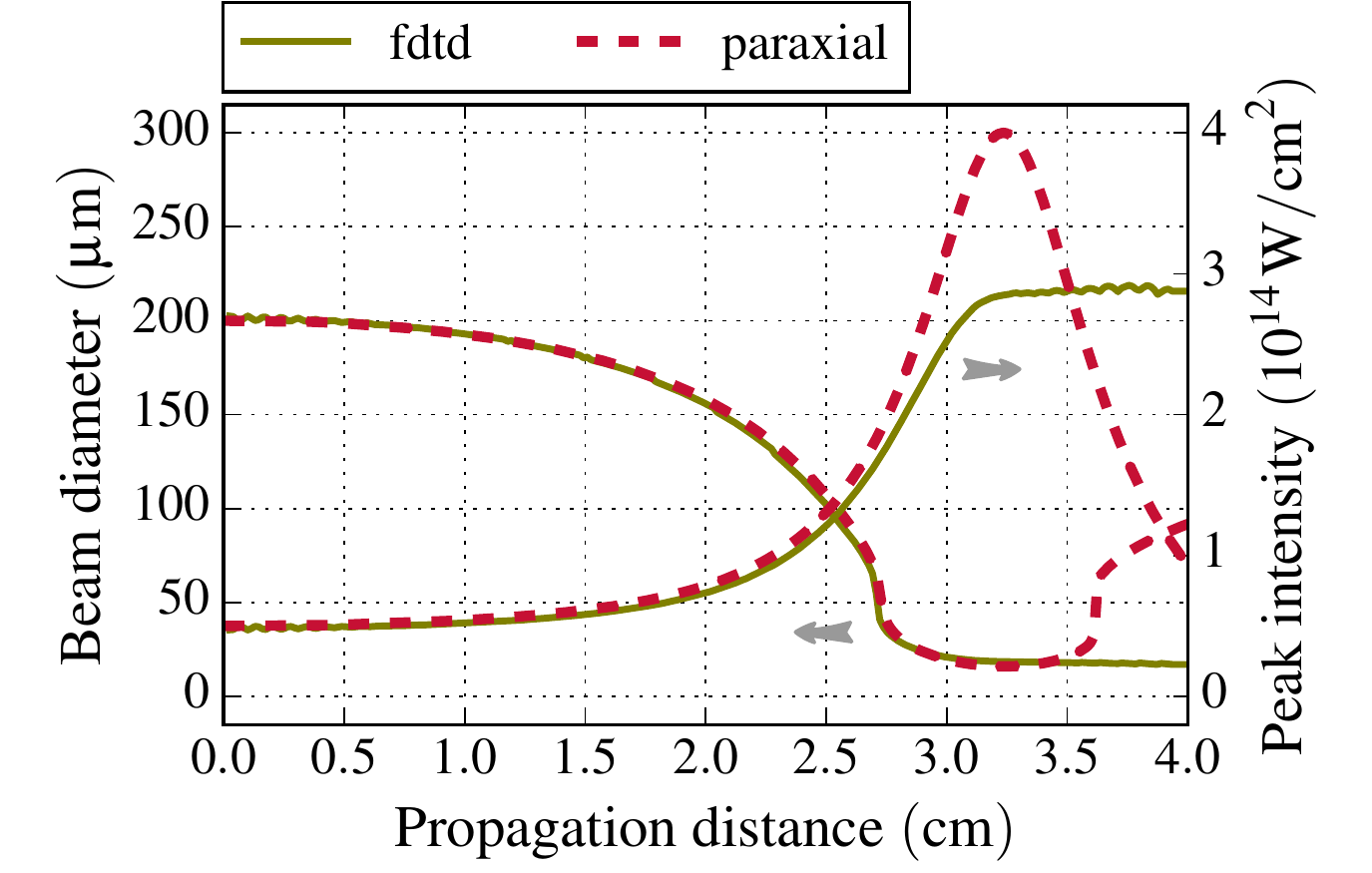}
\caption{Beam diameter and peak intensity of an 25-fs 800-nm laser pulse propagating in a fictitious gas medium modelled with a single strong-field oscillator equation like Eq.~\eqref{eq:nonlinear_lorentz_sat2}. Parameters are $\omega_0=2.9\times 10^{16}\,\mathrm{rad/s}$, $\gamma = 0$, $\bar{\chi}^{(1)}=6\times 10^{-4}$, and $\bar{\chi}^{(3)}=1.602\times 10^{-25}\,\mathrm{m}^2/\mathrm{W}$, which gives an optical response comparable to that of air at standard ambient temperature and pressure. Initial laser intensity was set to a high value for self-focusing to occur within a short distance to reduce the computation time.}
\label{fig:self_focusing}
\end{figure*}

\begin{figure*}[]
\centering
\includegraphics[width=1.0\columnwidth]{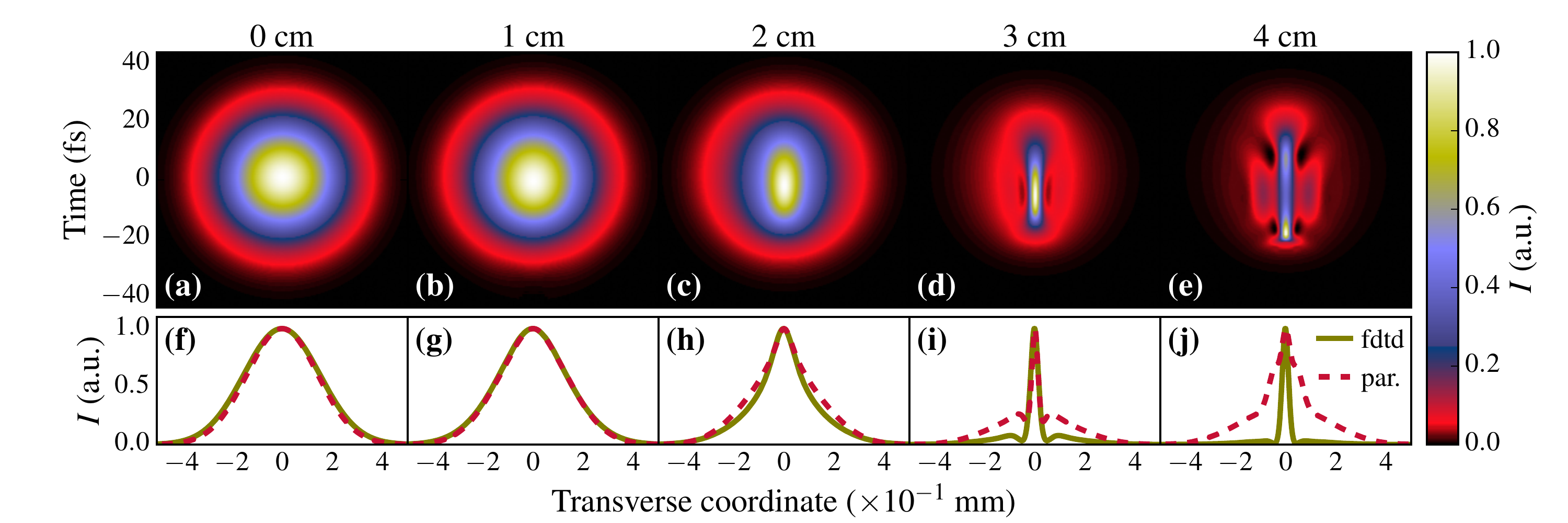}
\caption{Normalized electric field intensity ($\propto |\mathbf{E}|^2$) snapshots of self-focusing of the optical pulse of Fig.~\ref{fig:self_focusing}. Top row are FDTD results at different propagation distances. Bottom are transverse cuts of the top images along the plane of the peak intensity compared with results obtained with the paraxial equation given in \ref{apppendix:paraxial} (referred to as ``par.'' in the legend). All intensities are normalized to be directly comparable (corresponding values are found in Fig.~\ref{fig:self_focusing}). We see that at the early stage of the self-focusing process [(a),(b),(f), and (g)], the pulse remains symmetric and that the FDTD and paraxial models are in perfect agreement. But as the pulse collapses, it starts splitting longitudinally [see (d) and (e)], which stops prematurely the increase of the peak intensity, as observed in Fig.~\ref{fig:self_focusing}. In (i) and (j), it is also observed that the longitudinal splitting is coupled with important changes of the transverse profile, as compared with the paraxial model.}
\label{fig:self_focusing_frames}
\end{figure*}

From both Figs.~\ref{fig:strong_field} and \ref{fig:strong_field_I_scan} it appears that the anharmonic oscillator model is numerically unstable and it is not recommended for strong-field simulations. We stress that this numerical instability was also pointed out by independent authors (see, in particular,~\cite{Gordon2013,Scalora2015}). We emphasize that in the strong-field regime, it effectively fails at reproducing any of the two-level or instantaneous Kerr polarizations and their respective intensity-scaling behaviour. It is shown in \ref{appendix:A} that the anharmonic oscillator can also be derived from the two-level atom model equations. However, it stands on three important approximations: 1. the electric field is weak enough so that the nonlinearity can be treated as a small perturbation, 2. there is no population dynamics ($w\simeq -1$ and $dw/dt\simeq 0$), and 3. the nonlinear terms proportional to $\mathbf{P}^n$ with $n > 3$ are neglected (quadratic restoring force approximation). We identify that in the strong field limit the first and second approximations lead to inaccurate predictions of the optical nonlinearity even for moderate field strengths (see, e.g.,~\cite{VarinOE2015}), while the third one leads to numerical instability, in agreement with observations reported in~\cite{Gordon2013}. Our analysis thus suggests that the anharmonic (Duffing) oscillator [Eq.~\eqref{eq:anharmonic_oscillator}] should not be used for modeling nonlinear optical effects in moderate to intense fields with time-domain techniques like FDTD (see also~\cite{VarinOE2015}).

Ultimately, we modelled self-focusing of an optical pulse in a fictitious gas with Eq.~\eqref{eq:nonlinear_lorentz_sat2}. Two-dimensional FDTD simulations were compared against numerical results obtained with the nonlinear paraxial wave equation given in \ref{apppendix:paraxial}. In Fig.~\ref{fig:self_focusing}, it is effectively observed that at the beginning of the self-focusing process both the strong-field-Lorentz/FDTD and paraxial models agree perfectly. But as the beam collapses onto itself, significant differences arise. Effectively, it is shown that the paraxial equation predicts self-focusing of the pulse, followed almost immediately by defocusing. This is led by diffraction that then overcomes the self-focusing effect (which will eventually take over again to refocus, and so on). However, the 2D-FDTD results suggest a different scenario, where both the peak intensity and beam diameter level off and stay constant, at least for the propagation distance we covered. The smallest spot size observed in Fig.~\ref{fig:self_focusing} is approximately 16~$\mu$m ($\sim 20\lambda_L$). It is thus safe to assume that the paraxial approximation is not at the origin of the difference between the two models. Effectively, the sample snapshots of the self-focusing optical pulse given in Fig.~\ref{fig:self_focusing_frames} show that the finite duration of the pulse, although not so short (initially $\approx 25\,\mathrm{fs}$), plays a key role. The disagreement is thus better explained by longitudinal effects that are not considered in the nonlinear paraxial model. 

We emphasize that it is well known that time dependence has to be included when dealing with optical pulses, in particular in the context of self-focusing and laser filamentation~\cite{Berge2007,couaironPR2007}. In that respect, the time-independent paraxial model is not ideal. We stress that we used it here for validation purposes. This is fully justified, as in Figs.~\ref{fig:self_focusing} and~\ref{fig:self_focusing_frames} there is a near-perfect agreement between the paraxial and nonlinear-Lorentz/FDTD models up to the collapse ($\sim 2.75\,\mathrm{cm}$). This demonstrates unambiguously the validity of the strong-field oscillator model of Eq.~\eqref{eq:nonlinear_lorentz_sat2} and its numerical stability in situations where the numerical integration of the anharmonic (Duffing) oscillator does not converge.

With this last example, we have shown again that the nonlinear-Lorentz method is a rigorous and transparent approach to the modelling of linear and nonlinear optical phenomena in the FDTD framework. Moreover, it provides both qualitative and quantitative insight in situations where simpler, approximate models miss important physical contributions. Complexity could have certainly been added to the paraxial equation, e.g., finite pulse duration and spectral shape, dispersion, vectorial treatment, non-paraxiality, ... but all these contribution are intrinsic to the nonlinear-Lorentz/FDTD model, in particular here in its strong-field formulation.


\section{Conclusions}
\label{sec:conclusions}

In this paper, we have presented an extension of the Lorentz dispersion model to include nonlinear optics in the FDTD framework. Compared with the current iterative techniques, its numerical integration is simple, intuitive, fully explicit, and flexible. A complete methodology was elaborated, supported by full-scale FDTD tests that were compared with known and accepted theoretical models and calculations. We have shown that three-dimensional FDTD implementation is straightforward due to the inherent vectorial nature of the model, with the possibility to include damping, restoring, and light-matter coupling in a tensorial form to allow effective modeling of anisotropic nonlinear response. Finally, we proposed and analyzed a formulation of the nonlinear Lorentz model suitable for strong field applications. The current work suggests that it should be used in replacement of the anharmonic (Duffing) oscillator that tends to be unstable and inaccurate for moderate to intense fields. Ultimately, the use of the strong-field nonlinear-Lorentz model in FDTD-based plasma simulation approaches like PIC and MicPIC promises insight into the light-matter interaction processes involved in situations where both the dielectric polarization and plasma dynamics are important, e.g., during laser filamentation and femtosecond micromachining of dielectrics.

\vspace{\baselineskip}
\textbf{Acknowledgments.}  
The authors thank the EPOCH development and support team for their precious help. 
C.V. thanks Antonino Cal\`a Lesina for stimulating discussions. This research was supported by the Natural Sciences and Engineering Research Council of Canada (NSERC) through the Canada Research Chair in Ultrafast Photonics. EPOCH development was funded by the UK EPSRC grants EP/G054950/1, EP/G056803/1, EP/G055165/1 and EP/ M022463/1. Computations were made on the supercomputer Briar\'ee from Universit\'e de Montr\'eal, managed by Calcul Qu\'ebec and Compute Canada. The operation of this supercomputer is funded by the Canada Foundation for Innovation (CFI), the Minist\`ere de l'\'Economie, de la Science et de l'Innovation du Qu\'ebec (MESI) and the Fonds de recherche du Qu\'ebec - Nature et technologies (FRQ-NT). T.F. acknowledges financial support by the German Science Foundation (DFG) via SFB652/3 and SPP1840.

\appendix
\section{Formal derivation of the nonlinear Lorentz dispersion model}\label{appendix:A}

Atomic optical response in the weak-field, under-resonant limit can be approximated as the summation over individual, uncorrelated two-level transitions. This provides a good approximation to a more complete theory of the optical polarizability of atoms~(see, e.g.,~\cite{bethe1957quantum}). The optical response of a single transition is then conveniently modelled by the quantum mechanical two-level atom model whose harmonic oscillator formulation is characterized by the following coupled equations~\cite{boyd2008nonlinear,siegman1986lasers}:
\begin{align}\label{eq:harmonic_two_level_a}
\frac{d^2 \mathbf{p}}{dt^2} + \frac{2}{T_2} \frac{d \mathbf{p}}{dt} +  \omega_{0}^2\mathbf{p} &= -\kappa w\mathbf{E}\\ 
\frac{d w}{dt} + \frac{w - w^{eq}}{T_1} &= \left(\frac{2}{\hbar\omega_{0}}\right)\mathbf{E}\cdot \frac{d \mathbf{p}}{dt}.
\label{eq:harmonic_two_level_b}
\end{align}
$\mathbf{E}$ is the electric field vector of the driving signal, $\mathbf{p}$ is the expectation value of the induced dipole moment, $w$ is the population inversion parameter with equilibrium value $w^{eq}$, $\kappa = 2\omega_0\mu_0^2/\hbar$ is an atom-field coupling parameter, $\mu_0$ is the atomic dipole transition constant, $\hbar\omega_0$ is the transition energy between the two levels, $T_1$ is the population relaxation time, and $T_2$ is the atomic dephasing time. 

In the linear limit where $w\simeq -1$, Eqs.~(\ref{eq:harmonic_two_level_a}) and~(\ref{eq:harmonic_two_level_b}) simplify to the Lorentz dispersion model given at Eq.~\eqref{eq:lorentz_drude}. Below, we look at the perturbative nonlinear regime to find an approximate solution for the population parameter $w(t)$. When this approximate solution is reintroduced in Eq.~\eqref{eq:harmonic_two_level_a} and Taylor expanded in powers of $\mathbf{E}$ around $\mathbf{E}=0$, the nonlinear Lorentz model equation [Eq.~\eqref{eq:nonlinear_lorentz_drude}] is found.

We now consider an electromagnetic pulse whose electric field is defined by $\mathbf{E}(t) = \mathbf{E}_0\sin\left(\omega_L t\right)\exp(-t^2/t_p^2)$, with amplitude $\mathbf{E}_0$, angular frequency $\omega_L$, and duration $t_p$. Then, we assume that the atom-light interaction is under-resonant ($\omega_L \ll \omega_0$). Next, we consider the adiabatic-following approximation by assuming that the duration of the pulse is short enough so that we can neglect the two relaxation constants ($t_p \ll T_1, T_2$) by taking $T_2\rightarrow\infty$ and $T_1\rightarrow\infty$ in Eqs.~(\ref{eq:harmonic_two_level_a}) and~(\ref{eq:harmonic_two_level_b}), respectively. Under these approximations, $\mathbf{p} \simeq -\kappa w\mathbf{E}/\omega_{0}^2$ and
\begin{align}\label{eq:w_under_adiabatic}
\frac{d w}{dt} \simeq  -\left(\frac{2\kappa}{\hbar\omega_{0}^3}\right)\left[|\mathbf{E}|^2 \frac{d w}{dt} + \frac{w}{2} \frac{d |\mathbf{E}|^2}{dt}\right],
\end{align}
where $|\mathbf{E}|^2 = \mathbf{E}\cdot \mathbf{E}$ is a quantity oscillating at twice the laser frequency $\omega$, in comparison with the envelope (associated with $\mathbf{E}\cdot \mathbf{E}^*$). Eq.~\eqref{eq:w_under_adiabatic} has the exact solution that follows:
\begin{align}\label{eq:w_under_adiabatic_sol}
w = \frac{1}{\sqrt{1 + \left(\frac{2\kappa}{\hbar\omega_{0}^3}\right)|\mathbf{E}|^2}} + C,
\end{align}
in agreement with \cite{boyd2008nonlinear}. In the limit $t\rightarrow -\infty$, $\mathbf{E} = 0$ and $w = -1$ (the population is in the ground state), leading to $C = -2$. Replacing $w$ in Eq.~\eqref{eq:harmonic_two_level_a} by Eq.~\eqref{eq:w_under_adiabatic_sol} then gives:
\begin{align}\label{eq:nonlinear_lorentz}
\frac{d^2 \mathbf{p}}{dt^2} + \frac{2}{T_2} \frac{d \mathbf{p}}{dt} +  \omega_{0}^2\mathbf{p} &= \kappa \left(2- \frac{1}{\sqrt{1 + \left(\frac{2\kappa}{\hbar\omega_{0}^3}\right)|\mathbf{E}|^2}}\right)\mathbf{E}.
\end{align}
This is a quantum-mechanically valid form of the two-level atom model in the under-resonant, adiabatic-following limit. The validity conditions of this equation are typically met for moderately intense femtosecond pulses in dielectrics, where $T_1$ and $T_2$ have typical values in the 1-100~ns and 1-100~ps range, respectively. 

Expanding the square root in powers of $\left(2\kappa/\hbar\omega_{0}^3\right)|\mathbf{E}|^2$ around $|\mathbf{E}| = 0$ further gives:
\begin{align}\label{eq:nonlinear_lorentz_perturbative}
\frac{d^2 \mathbf{p}}{dt^2} + \frac{2}{T_2} \frac{d \mathbf{p}}{dt} +  \omega_{0}^2\mathbf{p} =  \kappa \mathbf{E} + \frac{\kappa}{2} \left(\frac{2\kappa}{\hbar\omega_{0}^3}\right)\mathbf{E}^3 - \frac{3\kappa}{8} \left(\frac{2\kappa}{\hbar\omega_{0}^3}\right)^2\mathbf{E}^5 + \ldots
\end{align}
It is then interesting to develop Eq.~\eqref{eq:nonlinear_lorentz_perturbative} by inserting the linearized ($w\simeq -1$) under-resonant adiabatic-following relationship $\mathbf{p} \simeq -\kappa w\mathbf{E}/\omega_{0}^2\simeq \kappa \mathbf{E}/\omega_{0}^2$ into the nonlinear source terms. 
After simplifying and rearranging terms:
\begin{align}\label{eq:anharmonic_oscillator}
\frac{d^2 \mathbf{p}}{dt^2} + \frac{2}{T_2} \frac{d \mathbf{p}}{dt} +  \omega_{0}^2\mathbf{p} - \frac{1}{2}\left(\frac{\omega_0}{\mu_0}\right)^2\mathbf{p}^3 + \mathcal{O}(\mathbf{p}^5) - \ldots = \kappa \mathbf{E}.
\end{align}
Eq.~\eqref{eq:anharmonic_oscillator} is identified as the anharmonic (Duffing) oscillator equation with a third-order nonlinear parameter $b = \omega_0^2/2\mu_0^2$ (see also \cite{Gordon2013,Scalora2015}). It is then straightforward to show that Eq.~\eqref{eq:anharmonic_oscillator} reduces to the linear, Lorentz dispersion equation when $|\mathbf{p}|\ll \sqrt{2}\mu_0$, where, again, $\mu_0$ is the atomic dipole moment constant associated with the strength of the atomic transition.

We stress that the right-hand side of Eq.~\eqref{eq:nonlinear_lorentz_perturbative} displays odd powers of $\mathbf{E}$, providing modelling capabilities only for centrosymmetric material. In certain dielectrics, the unit crystal cell is associated with constant crystal fields that break the centrosymmetry of the wave function and give rise to even-order contributions. To apply the two-level atom model to the modelling of non-centrosymmetric media, we extend the derivation provided above to the situation where both the ground state and the excited state have a permanent dipole moment. Starting from the density matrix formulation of the two-level atom, this is done by assigning non-zero values to the diagonal elements of the dipole moment operator matrix. Following the procedure found in Sec.~6.4.1 in~\cite{boyd2008nonlinear}, it is possible to show that the polarization and population equations are then
\begin{align}\label{eq:p}
	\frac{d^2 p}{dt^2} + \frac{2}{T_2}\frac{d p}{dt}+\omega_{0}^2\left(1 
	+\frac{\Delta\mu}{\hbar\omega_{0}} E\right)p
	&=\omega_{0}^2\left(1 
	+\frac{\Delta\mu}{\hbar\omega_{0}} E\right)\left(\bar{\mu} - \frac{w}{2}\Delta\mu\right)
	-\kappa Ew\\
	\frac{d w}{dt} + \frac{w - w^{(\mathrm{eq})}}{T_1} &= \left(\frac{2}{\hbar\omega_{0}}\right)E\,\frac{d p}{dt}\label{eq:w}
\end{align}
where $\Delta \mu = \mu_g - \mu_e$ is the difference between the permanent dipole moments of the ground ($\mu_g$) and excited ($\mu_e$) states, and $\bar{\mu} = (\mu_g + \mu_e)/2$. The other parameters are as defined in the text below Eqs.~\eqref{eq:harmonic_two_level_a} and~\eqref{eq:harmonic_two_level_b}.

In the adiabatic-following, under-resonant limit ($\omega_L \ll \omega_0$ and $T_2\rightarrow\infty$):
\begin{align}\label{eq:p_approx}
p \simeq \left(\bar{\mu}-\frac{w}{2}\Delta\mu\right) -\frac{\kappa wE}{\omega_{0}^2\left(1+\frac{\Delta \mu}{\hbar\omega_0}E\right)}.
\end{align}
Inserting Eq.~\eqref{eq:p_approx} into Eq.~\eqref{eq:w} (with $T_1\rightarrow\infty$) leads to
\begin{align}\label{eq:dw}
\frac{dw}{w} &= -\frac{bE}{\left[(1 + aE)^2 + bE^2\right]\left(1 +a E\right)}\,dE,
\end{align}
where $a=\Delta \mu/\hbar\omega_0$ and $b=2\kappa/\hbar\omega_{0}^3$. The exact analytical solution to Eq.~\eqref{eq:dw} is
\begin{align}\label{eq:w_sol}
w &= \frac{1 + aE}{\left[(1 + aE)^2 + bE^2\right]^{1/2}} + C.
\end{align}
We know from the previous solution given in Eq.~\eqref{eq:w_under_adiabatic_sol} that $C = -2$, such that $w = -1$ when $E = 0$. Moreover, we emphasize that Eq.~\eqref{eq:w_under_adiabatic_sol} is recovered with Eq.~\eqref{eq:w_sol} when $a=0$, i.e., when there are no permanent dipole moments associated with the ground state and excited state.

By expanding Eq.~\eqref{eq:w_sol} in a Taylor series around $E = 0$, it is shown that the weak-field form of Eq.~\eqref{eq:p} is
\begin{align}\label{eq:harmonic_with_even}
\frac{d^2 p}{dt^2} + \frac{2}{T_2}\frac{d p}{dt}+\omega_{0}^2p
=\omega_0^2\left\lbrace \mu_{g} + \left(c + a\bar{\mu}\right)E+\frac{bd}{2}E^2+\frac{\left(bc - 2abd\right)}{2}E^3+\ldots\right\rbrace,
\end{align}
where $a$ and $b$ are defined below Eq.~\eqref{eq:dw}, $c = a\Delta\mu/2 + b$, and $d = \Delta\mu/2$. We emphasize that the first term on the right-hand side of Eq.~\eqref{eq:harmonic_with_even} is the permanent dipole moment of the ground state. Since it is constant, it is usually neglected in an FDTD treatment. Then Eq.~\eqref{eq:harmonic_with_even} takes the following form:
\begin{align}\label{eq:harmonic_with_even_chi}
\frac{d^2 p}{dt^2} + \frac{2}{T_2}\frac{d p}{dt}+\omega_{0}^2p
=\omega_0^2\epsilon_0\left\lbrace \bar{\chi}^{(1)}E+\bar{\chi}^{(2)}E^2+\bar{\chi}^{(3)}E^3+\ldots\right\rbrace,
\end{align}
where $\bar{\chi}^{(1)} = \left(c + a\bar{\mu}\right)/\epsilon_0$, $\bar{\chi}^{(2)} = bd/2\epsilon_0$, $\bar{\chi}^{(3)} = \left(bc - 2abd\right)/2\epsilon_0$, $\ldots$ 
This demonstrates formally that Eq.~\eqref{eq:nonlinear_lorentz_drude} presented in Sec.~\ref{sec:nonlinear}~is a legitimate, macroscopic generalization of the quantum mechanical two-level atom model for both centrosymmetric and non-centrosymmetric media in the weak-field ($|\mathbf{E}|\rightarrow 0$) under-resonant ($\omega_L \ll \omega_0$) adiabatic-following ($t_p \ll T_1, T_2$) limit.

Deeper insight into the weak-field condition ($|\mathbf{E}|\rightarrow 0$) used above is gained if we compare the proper light-matter interaction timescale (say $T$) with the Rabi frequency $2\mu_0|E|/\hbar$. Effectively, a laser field will be considered weak if the Rabi oscillations are slow compared with $T$, i.e., if the Rabi phase is nearly constant over the timescale $T$. This happens when $2\pi/T \gg 2\mu_0|E|/\hbar$ (see also Sec.~6.4.2 in \cite{boyd2008nonlinear}). This allows to define an upper limit for the laser intensity as $I^* = |E|^2/2\eta_0 = (\pi\hbar/T\mu_0)^2/2\eta_0$. Typically, $\mu_0$ is on the order of one atomic unit ($\mu_0\simeq 8.478\times 10^{-30}\,\mathrm{Cm}$). For the interaction of short pulses with dielectrics, the relevant timescale $T$ is the pulse duration. For a 15-fs pulse, $I^* \sim 10^{12}\,\mathrm{W/cm}^2$. This shows that a typical upper laser intensity limit to respect the ``weak-field'' condition of the nonlinear Lorentz model is actually quite high: above the typical values encountered in perturbative nonlinear optics and in a range where avalanche breakdown can be triggered in dielectrics.

\section{Range of validity of the nonlinear Lorentz dispersion model with respect to the laser frequency}\label{appendix:validity}
In~\ref{appendix:A}, we have shown how the nonlinear Lorentz dispersion model is obtained from the quantum-mechanical two-level atom model in the weak-field adiabatic-following under-resonant limit. To test the validity range with respect to the laser frequency $\omega_L$, we performed a series of simulations identical to that of Figs.~\ref{fig:strong_field} and~\ref{fig:strong_field_I_scan} and scanned $\omega_L$ from the under-resonant limit ($\omega_L < \omega_0$) to the over-resonant limit ($\omega_L > \omega_0$),  where $\omega_0$ is the oscillator natural frequency. Dispersion curves for both the first-order ($|P(\omega_L)|$) and third-harmonic ($|P(3\omega_L)|$) polarization densities were then obtained. We used the numerical solution of the quantum-mechanical two-level atom equations as a reference (see~\ref{appendix:schrodinger} for details).

To support the analysis that follows, we recall that the anharmonic oscillator Eq.~\eqref{eq:anharmonic_oscillator} has a perturbative solution (see, e.g.,~\cite{boyd2008nonlinear,Scalora2015}) that we write in the following form (note the neglect of damping, for simplicity):
\begin{subequations}
	\begin{align}\label{eq:P1}
	P(\omega_L) &= \frac{\epsilon_0\bar{\chi}^{(1)}}{\left(1-\frac{\omega_L^2}{\omega_0^2}\right)}E_0\\
	P(3\omega_L) &= \frac{\epsilon_0\bar{\chi}^{(3)}}{\left(1-\frac{\omega_L^2}{\omega_0^2}\right)^3\left(1-\frac{9\omega_L^2}{\omega_0^2}\right)}E_0^3,\label{eq:P3}
	\end{align}
\end{subequations}
where $E_0$ is the amplitude of the driving electric field (monochromatic) and $\bar{\chi}^{(1)}$ and $\bar{\chi}^{(3)}$ are the first-order and third-order static susceptibility parameters, respectively. The dispersion curve for the third harmonic is thus expected to show two distinct resonance peaks: one at $\omega_L = \omega_0$ and another at a lower frequency $\omega_L = \omega_0/3$. For a second-order material, the second peak appears effectively at $\omega_L = \omega_0/2$, whereas the first peak position at $\omega_L = \omega_0$ is unchanged.

Results shown in Fig.~\ref{fig:laser_freq_scan} confirm that linear dispersion is correctly modelled by the different polarization equations presented above in Table~\ref{table:strongfield}. Effectively, curves in Fig.~\ref{fig:laser_freq_scan}(a) are barely distinguishable over the entire frequency range we considered ($0.065 \leq \omega_L/\omega_0 \leq 2.2$). For third-harmonic dispersion, agreement is good up to the first resonance of Eq.~\eqref{eq:P3} ($\omega_L = \omega_0/3$), which marks a tangible upper limit to the laser angular frequency. Beyond that mark and through the main resonance ($\omega_L = \omega_0$), none of the approximate solutions match the quantum mechanical prediction. This emphasizes that all three models (nonlinear Lorentz, anharmonic oscillator, and strong-field Lorentz) are in fact special-case equations that are valid in the under-resonant frequency limit only. The current analysis shows that the turning point is $\omega_L \simeq \omega_0/3$ (again, $\omega_0/2$ for a second-order material).

\begin{figure*}[ht]
	\centering
	\includegraphics[width=\columnwidth]{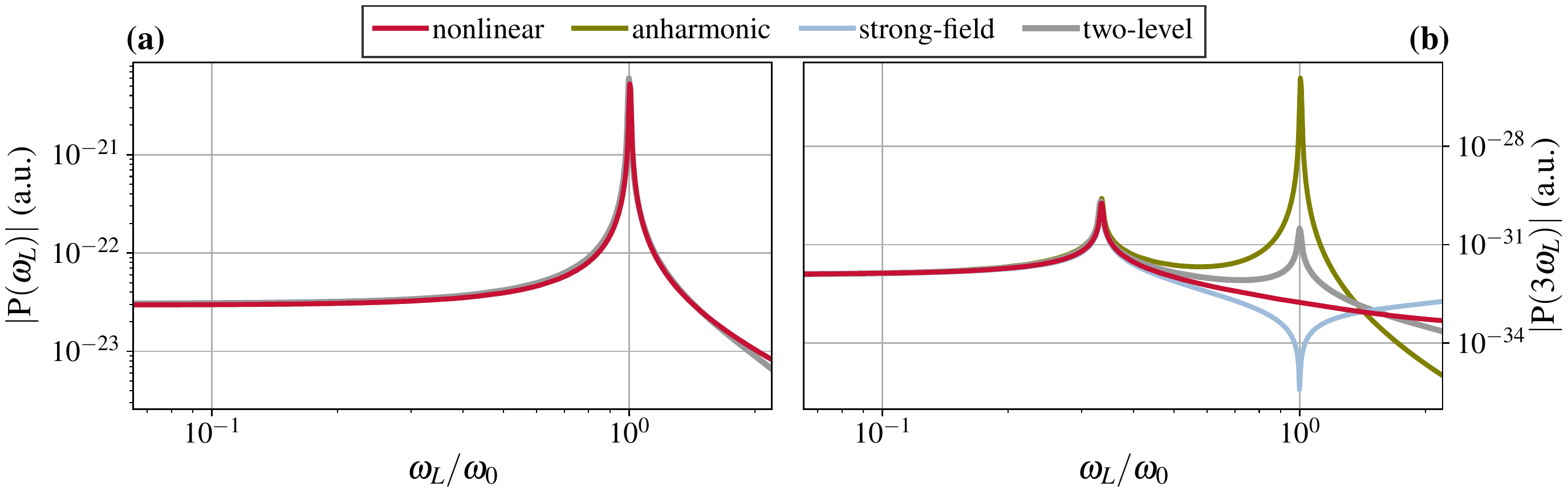}
	\caption{Dispersion curves for (a) linear and (b) third-harmonic scattering as predicted by the different nonlinear oscillator equations presented in Sec.~\ref{sec:strong} (see, in particular, Table~\ref{table:strongfield} and caption of Fig.~\ref{fig:strong_field} for details). While the difference between all models in (a) is barely distinguishable, agreement in (b) is limited to the under-resonant range ($\omega_L \lesssim \omega_0/3$). Further analysis is provided in Fig.~\ref{fig:laser_freq_scan_error}.
	}
	\label{fig:laser_freq_scan}
\end{figure*}

\begin{figure*}[ht]
	\centering
	\includegraphics[width=\columnwidth]{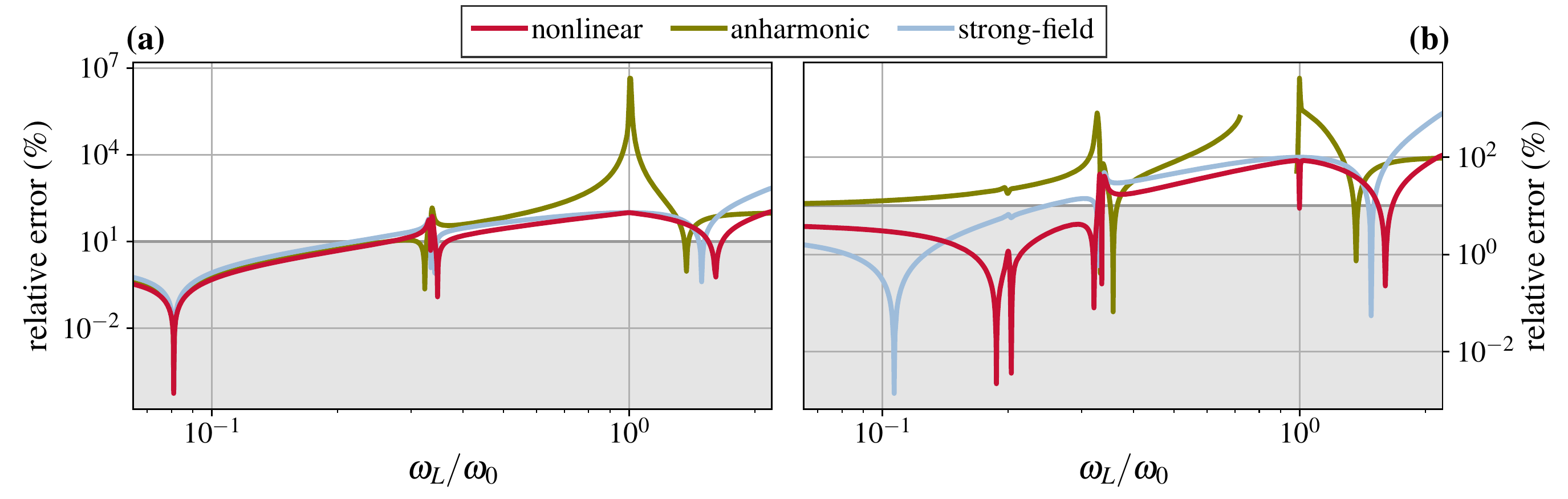}
	\caption{Error analysis of the third-harmonic dispersion curves presented in Fig.~\ref{fig:laser_freq_scan}. The relative error is defined as $100\times|model - quantum|/quantum$, where $model$ and $quantum$ are the predictions from a specific model (see legend) and that of the quantum-mechanical two-level atom model, respectively (see caption of Fig.~\ref{fig:strong_field} for details). The grey areas are regions where the error is under 10\%. In (a), for a weak laser intensity ($10^6\,\mathrm{W/cm}^2$), the error is effectively confined within a 10\% margin 
	for $\omega_L \lesssim 0.25 \omega_0$, which is slightly narrower than what was inferred from Eq.~\eqref{eq:P3} and Fig.~\ref{fig:laser_freq_scan}(b) ($\omega_L \leq \omega_0/3$). With increasing laser intensity, the error associated with the anharmonic oscillator model steadily moved away from the grey region, while that associated with the strong-field Lorentz model did not change significantly. In particular for the nonlinear Lorentz model, the below-10\% range got wider with increasing laser intensity. In (b), it is seen that at a $10^{13}\,\mathrm{W/cm}^2$ laser intensity it extends all the way up to the $\omega_L = \omega_0/3$ turning point.
	}
	\label{fig:laser_freq_scan_error}
\end{figure*}

Better insight is gained by plotting the relative error of the different model predictions with respect to the quantum solution (see Fig.~\ref{fig:laser_freq_scan_error}). At low intensity ($10^6\,\mathrm{W/cm}^2$), the three models match the quantum solution within a 10\% error up to $\omega_L \simeq 0.25\omega_0$. With increasing laser intensity, the error associated with the nonlinear and strong-field Lorentz models stays in the same range, whereas that associated with the anharmonic oscillator model increased steadily [see also Fig.~\ref{fig:strong_field_I_scan}(a) for details about the intensity scaling of the different models]. In particular, it is seen in Fig.~\ref{fig:laser_freq_scan_error}(b) that for high laser intensity ($10^{13}\,\mathrm{W/cm}^2$) the error for the nonlinear Lorentz model is kept within 10\% over the range $0.065 \leq \omega_L/\omega_0 \leq 0.33$, while in the same range the anharmonic solution error is systematically greater than 10\%. Finally, we observed that laser intensity has a minor influence on the error associated with the strong-field Lorentz model.

In Secs.~\ref{sec:example1} and~\ref{sec:3D}, we presented examples involving second-order materials. The oscillator frequencies we used for the underlying nonlinear Lorentz equation were $\omega_0 = 1.55\times 10^{16}\,\mathrm{rad/s}$ and $\omega_0 = 5.18\times 10^{15}\,\mathrm{rad/s}$, for laser wavelengths of $\lambda_L = 1.064~\mu\mathrm{m}$ and $\lambda_L = 1.5~\mu\mathrm{m}$, respectively. Frequency ratios are then found to be $\omega_L/\omega_0 \simeq 0.114$ and $0.251$, comfortably within the validity range for second-order material ($\omega_L \lesssim \omega_0/2$). For the examples with third-order dielectrics in Secs.~\ref{sec:example2} ($\lambda_L = 1.5~\mu\mathrm{m}$, $\omega_0 = 2.75\times 10^{16}\,\mathrm{rad/s}$) and~\ref{sec:strong} ($\lambda_L = 800~\mathrm{nm}$, $\omega_0 = 3\times 10^{16}\,\mathrm{rad/s}$), frequency ratios where $0.0457$ and $0.0785$. Comparison of these ratios with Fig.~\ref{fig:laser_freq_scan_error} suggests that modelling the medium with the nonlinear and strong-field Lorentz equations introduced an error below 10\% relative to modelling with the quantum-mechanical two-level atom model.

\section{Nonlinear-Lorentz modelling of optical dispersion}\label{appendix:dispersion}
%

Direct access to the spectral contributions associated with the static susceptibility parameters $\bar{\chi}^{(\xi)}$ of the nonlinear Lorentz model is obtained by taking the Fourier transform $\mathrm{FT}\{\quad\}$ of Eq.~\eqref{eq:nonlinear_lorentz_drude}, which gives:
\begin{align}\label{eq:fourier}
\tilde{\mathbf{P}}(\omega) = \epsilon_0\left(\frac{\omega_0^2}{\omega_0^2 - \omega^2 - i\gamma\omega}\right)\sum_{\xi}\bar{\chi}^{(\xi)}\, \mathrm{FT}\{\mathbf{E}^\xi\}.
\end{align}
It is seen immediately that $\tilde{\mathbf{P}}(\omega)$ is the sum over all the $\xi$th-order terms that obey:
\begin{align}\label{eq:nth_order_fourier}
\tilde{\mathbf{P}}^{(\xi)}(\omega) = \epsilon_0\left(\frac{\omega_0^2}{\omega_0^2 - \omega^2 - i\gamma\omega}\right)\bar{\chi}^{(\xi)}\, \mathrm{FT}\{\mathbf{E}^\xi\}.
\end{align}
This natural spectral decomposition is equivalent to defining an independent oscillator equation for every order $\xi$ as
\begin{align}\label{eq:nth_order_equation}
\frac{d^2 \mathbf{P}^{(\xi)}}{d t^2} + \gamma\frac{d \mathbf{P}^{(\xi)}}{d t} +  \omega_0^2\mathbf{P}^{(\xi)} & = \omega_0^2\epsilon_0\bar{\chi}^{(\xi)} \mathbf{E}^\xi.
\end{align}
We emphasize that $\mathbf{P}^{(\xi)}$ above does not oscillate only at the harmonic frequency $\xi\omega_L$, where $\omega_L$ is the laser frequency, as one would expect from a regular perturbative decomposition of $\mathbf{P}$ (see, e.g.,~\cite{boyd2008nonlinear}). We recall that $\mathbf{E}$ is here a real quantity that corresponds to the total electric field. Its $\xi$th power, i.e., $\mathbf{E}^\xi$, thus potentially contains several harmonic contributions that might overlap with those from the other terms, but uniquely scaled by the corresponding $\bar{\chi}^{(\xi)}$. For example, for a laser signal of the form $E = E_0\cos{\omega_L t}$, we can develop the first three orders according to Eq.~\eqref{eq:nth_order_fourier} to get:
\begin{align}
\tilde{P}^{(1)}(\omega) &=\epsilon_0\left(\frac{\omega_0^2}{\omega_0^2 - \omega^2 - i\gamma\omega}\right)\bar{\chi}^{(1)}\sqrt{\frac{\pi}{2}}\left[\delta(\omega - \omega_L)+\delta(\omega + \omega_L)\right]E_0.\\
\tilde{P}^{(2)}(\omega) &= \epsilon_0\left(\frac{\omega_0^2}{\omega_0^2 - \omega^2 - i\gamma\omega}\right)\bar{\chi}^{(2)}\sqrt{\frac{\pi}{2}}\left[\delta(\omega) + \frac{\delta(\omega - 2\omega_L)+\delta(\omega + 2\omega_L)}{2}\right]E_0^2\\
\tilde{P}^{(3)}(\omega) &= \epsilon_0\left(\frac{\omega_0^2}{\omega_0^2 - \omega^2 - i\gamma\omega}\right)\bar{\chi}^{(3)}\frac{\sqrt{2\pi}}{8}\nonumber\\
&\qquad\times\left[3\delta(\omega-\omega_L) + 3\delta(\omega+\omega_L) + \delta(\omega-3\omega_L) +\delta(\omega+3\omega_L)\right]E_0^3.
\end{align}
It is thus observed that the linear polarization term $\tilde{P}^{(1)}(\omega)$ peaks at $\pm\omega_L$, whereas the third order term has contributions at both $\pm\omega_L$ and $\pm3\omega_L$, associated with self-phase modulation and third harmonic generation, respectively. Also, $\tilde{P}^{(2)}(\omega)$ clearly exhibits the contributions associated with optical rectification [$\delta(\omega)$] and second harmonic generation [$\delta(\omega\pm 2\omega_L)$]. The strength of these linear and nonlinear effects are effectively weighted by the dispersion of the harmonic oscillator $\omega_0^2/(\omega_0^2 - \omega^2 - i\gamma\omega)$ and by the corresponding susceptibility parameters $\bar{\chi}^{(1)}$, $\bar{\chi}^{(2)}$, and $\bar{\chi}^{(3)}$.

In real materials, the spectral response of the various polarization orders is usually different. It is possible to account for this fact phenomenologically by using different oscillator parameters for the various polarizations orders. Also, by summing up the contributions from an ensemble of oscillators---labeled by the subscript $k$---it is possible to get an effective $\xi$th-order dispersion equation as
\begin{align}\label{eq:nth_order_chi}
\chi^{(\xi)}(\omega) = \sum_k \bar{\chi}^{(\xi)}_k(\omega) = \sum_k \left(\frac{\omega_k^2}{\omega_k^2 - \omega^2 - i\gamma_k\omega}\right)\bar{\chi}^{(\xi)}_k,
\end{align}
that can be used to obtain complex dispersion curves over an extended spectral range.
We emphasize that the particular case where $\xi = 1$ and $\gamma_k=0$, Eq.~\eqref{eq:nth_order_chi} leads to the Sellmeier dispersion formula for the linear refractive index $n(\omega) = \sqrt{1 + \sum_k\chi^{(1)}_k(\omega)}$~\cite{Sellmeier}, already widely used to fit spectroscopic measurement data. The nonlinear Lorentz model provides a natural and intuitive extension of this approach to the nonlinear domain.

In practical situations where dispersion modelling does not need high accuracy over a large spectral bandwidth, it is advantageous to reduce the number of model oscillators as much as possible to improve the computational efficiency. In fact, in many situations there is no need for proper modelling of the \emph{nonlinear} dispersion. This is the case for the examples we presented in Secs.~\ref{sec:example1}-\ref{sec:strong}, where we assumed that the nonlinear effects associated with $\chi^{(2)}$ and $\chi^{(3)}$ share the same dispersion as one of the oscillators of the (linear) Sellmeier formula. When there is no significant overlap between the laser pulse spectrum and the oscillator resonance, it is observed that $\omega_k^2/(\omega_k^2 - \omega^2)\simeq 1$ and the static nonlinear susceptibility parameters $\bar{\chi}^{(\xi)}$ can be used as-is (see, in particular, Fig.~\ref{fig:LiNbO3}). However, in general cases, $\bar{\chi}^{(\xi)}$ might have to be rescaled to get the desired nonlinear effects.

\section{Complementary equations}

The nonlinear-Lorentz/FDTD modelling examples we provided in this paper were compared with accepted models whose underlying equations are reproduced below.

\subsection{Coupled amplitude equations for second-harmonic generation}\label{appendix:shg}

Accepted models for second-harmonic generation (SHG) in quasi-phase-matched (QPM) crystals can be found in~\cite{Fejer1992,Byer1997,boyd2008nonlinear}. In particular, we integrated numerically Eqs.~(2.7.10) and (2.7.11) in~\cite{boyd2008nonlinear}, i.e., 
\begin{align}
\frac{dA_1}{dz} &= \frac{2i\omega_1^2d_\mathrm{eff}}{k_1c^2}A_2A_1^*e^{-i\Delta k z}\\
\frac{dA_2}{dz} &= \frac{i\omega_2^2d_\mathrm{eff}}{k_2c^2}A_1^2e^{i\Delta k z}.
\end{align}
Parameters are: $z$ the distance inside the crystal, $A_1$ the complex amplitude of the driving signal oscillating at $\omega_1$, $A_2$ the complex amplitude of the second harmonic oscillating at $\omega_2 = 2\omega_1$, $k_i=n(\omega_i)\omega_i/c$ are the corresponding wave-vectors that define the wave-vector mismatch parameter $\Delta k = 2k_1-k_2$, and, finally, $d_\mathrm{eff} = \bar{\chi}^{(2)}/2$. More details and derivations can be found in~\cite{boyd2008nonlinear}.

\subsection{Two-level nonlinear optics in the Schr\"odinger picture}\label{appendix:schrodinger}

In the Schr\"odinger picture, the dynamics of a two-level atom is studied in terms of the level probability amplitudes $C_a$ and $C_b$ whose temporal evolution is given by
\begin{align}
\frac{dC_a}{dt} &= \frac{1}{i\hbar}C_b\mu_{ab} E(t)e^{-i\omega_{ba}t}\\
\frac{dC_b}{dt} &= \frac{1}{i\hbar}C_a\mu_{ba} E(t)e^{i\omega_{ba}t},
\end{align}
where $\hbar \omega_{ba}$ is the transition energy and $\mu_{ij}$ are the transition matrix elements. These equations are essentially Eqs.~(6.5.6) and (6.5.8) found in~\cite{boyd2008nonlinear}. For the numerical integration we assumed a real electric field $E(t)$ and real matrix elements $\mu_{ab} = \mu_{ba} = \mu_0$. Then we evaluated the induced dipole moment as of Eq.~(6.5.31) in~\cite{boyd2008nonlinear}, i.e., 
\begin{align}
\langle\hat{\mu}\rangle = \mu_0\left(C_a^*C_be^{-i\omega_{ba}t} +C_aC_b^*e^{i\omega_{ba}t}\right).
\end{align}
Details and derivations are found in~\cite{boyd2008nonlinear}.

\subsection{Paraxial wave equation with Kerr nonlinearity}
\label{apppendix:paraxial}
In the lowest-order approximation, Kerr self-focusing is typically studied with the following nonlinear scalar paraxial wave equation, Eq.~(48) in \cite{Berge2007}, reproduced here:
\begin{align}
\frac{\partial U}{\partial z} = \frac{i}{2k_0}\nabla_\perp^2 U + i\frac{\omega_0}{c}n_2|U|^2U.
\end{align}
$U$ is the complex envelope, $k_0 = n_0\omega_0/c$ is the wave-vector associated with the linear index $n_0$ and angular frequency $\omega_0$, and $n_2$ is the third-order nonlinear index.

\vspace{\baselineskip}
\bibliography{references}

\end{document}